\documentclass[12pt,letterpaper]{article}
\usepackage[T1]{fontenc}
\usepackage[utf8]{inputenc}
\usepackage{lmodern,microtype}
\usepackage[margin=1.08in]{geometry}
\usepackage{amsmath,amssymb,amsthm,mathtools,booktabs,array,longtable}
\usepackage{setspace,enumitem,xcolor}
\usepackage[round,authoryear]{natbib}
\usepackage[colorlinks=true,linkcolor=blue,citecolor=blue,urlcolor=blue]{hyperref}
\setlist{itemsep=3pt,topsep=5pt,leftmargin=*}
\hypersetup{pdftitle={Quality at Stale Prices},pdfauthor={Georgy Lukyanov, Konstantin Shamruk, Ekaterina Logina}}
\numberwithin{equation}{section}
\newtheorem{theorem}{Theorem}[section]
\newtheorem{proposition}[theorem]{Proposition}
\newtheorem{lemma}[theorem]{Lemma}
\newtheorem{corollary}[theorem]{Corollary}
\newcommand{\E}{\mathbb E}\newcommand{\Prb}{\mathbb P}
\newcommand{\Abar}{\overline A}\newcommand{\Uti}{\widetilde U}
\newcommand{\ind}{\mathbf 1}
\newcommand{\argmax}{\operatorname*{arg\,max}}
\newcommand{\pos}[1]{[#1]_+}\newcommand{\Lip}{\operatorname{Lip}}

\title{Quality at Stale Prices\thanks{We thank Kirill Bukin, Harry Di Pei, Miaomiao Dong, Olivier Gossner, Christian Hellwig, Natalia Kovaleva, Yukio Koriyama, and participants at the 30th Stony Brook International Conference on Game Theory, the 2019 Econometric Society European Meeting (ESEM) in Manchester, and internal seminars at ICEF and NES for helpful comments. All remaining errors are our own.}}
\author{Georgy Lukyanov\thanks{Corresponding author. Toulouse School of Economics. Email: \href{mailto:georgy.lukyanov@tse-fr.eu}{georgy.lukyanov@tse-fr.eu}.}\and Konstantin Shamruk\thanks{Toulouse School of Economics. Email: \href{mailto:c.shamruk@gmail.com}{c.shamruk@gmail.com}.}\and Ekaterina Logina\thanks{International College of Economics and Finance, HSE University. Email: \href{mailto:loginka1995@gmail.com}{loginka1995@gmail.com}.}}
\date{October 14, 2026}
\begin{document}
\maketitle
\begin{abstract}
We study a continuous-time game in which a firm privately chooses quality and a public-information pricing desk revises its posted price at Poisson opportunities. Customers learn from purchases and silence. For an open set of primitives, we construct an equilibrium in which a previously optimal price sustains quality even though the desk would lower it upon revision. Quality costs exceed every immediate sales return, making the continuation reward necessary. We characterise both the resulting quality episodes and their lifetime incidence. A purchase ends an episode by raising reputation, a revision by changing the price, and silence by lowering reputation. Near the zero-quality threshold, the quality probability has a limiting parabolic profile and most episodes end through silence. If $\delta$ is the cost shortfall below that threshold, the expected lifetime number of high-quality decisions and the probability of any such decision are both of order $\delta^{3/2}$ from a specified upstream state. The persistent type is eventually learned, strategic quality ceases, and prices settle at type-specific levels. The construction is local at fixed finite discounting.
\end{abstract}
\clearpage

% BEGIN Introduction.tex
\section{Introduction}\label{sec:introduction}

A firm's reputation can change while its posted price remains in force. The firm may then alter the quality offered to a customer before it can revise the price. We ask whether a price inherited from an earlier, more favourable reputation can sustain costly quality after a lower price becomes optimal at a revision opportunity. We also ask how this incentive ends: through a purchase, through price adjustment, or through continued silence. These questions concern the same equilibrium path. Establishing that quality can occur does not by itself establish how much quality the firm will ever provide.

We consider a long-lived firm facing short-lived customers in continuous time. A committed type always supplies high quality; a strategic type chooses quality at each customer opportunity. Before buying, a customer receives a private signal of current quality. Purchases and posted prices are public, whereas quality, customer opportunities, and individual refusals are not. A pricing desk uses the public history to maximise posterior expected firm profit and can revise the price at independent Poisson opportunities. The strategic producer knows its type and controls quality, but cannot override the desk. This separation of information and decision rights is part of the pricing institution. The public probability of commitment and the inherited price jointly determine current incentives.

Our first result constructs an equilibrium in which the strategic firm supplies quality with positive probability at an inherited high price, although the desk would choose the low price upon revision. The quality cost exceeds every immediate sales return, so a continuation reward from a purchase is necessary. The high price can arise from an earlier optimal choice, and the quality region is reached with positive probability from that state. The construction is not confined to one parameter vector: its strict inequalities persist on an open set of primitives. Theorem~\ref{thm:primitive}, Corollary~\ref{cor:essential}, and Proposition~\ref{prop:robustness} establish these claims.

The distinction from the usual reputational return to quality is the disagreement between the inherited price and the current reset choice. Price affects both revenue and the sensitivity of demand to the customer's signal. A purchase improves the belief governing future demand, and supplying quality makes that purchase more likely. As reputation deteriorates, the price maximising the desk's value can therefore differ from the inherited price at which the strategic producer finds quality worthwhile. At a revision in the active region, the desk lowers the price and the quality incentive disappears. This is a comparison within one equilibrium, with the same belief and continuation strategies; it is not a comparison with a separately constructed flexible-price economy.

The equilibrium argument must account for the response of customers, beliefs, continuation values, and reset choices to anticipated quality. We begin from the zero-quality benchmark and lower cost just below its unrestricted deviation threshold. Quality then appears in a narrow belief region at the high price. Its small total mass makes the change in continuation values lower order than the direct cost change, allowing the inherited-price incentive to survive equilibrium feedback. The construction uses logistic demand, a two-price menu, and a sufficiently large finite discount rate fixed before the cost gap tends to zero. It establishes a local family of equilibria, not global existence or uniqueness for arbitrary primitives.

The adjustment results describe the extent of this same equilibrium response. The market eventually learns the persistent type; strategic quality ceases after a finite random time and prices settle at type-specific levels. Before that happens, a purchase can end a quality episode by raising reputation, a revision by changing the price, and silence by lowering reputation. If $\delta$ is the cost shortfall below the zero-quality threshold, the quality probability is of order $\delta$ over an interval of width $\sqrt\delta$, with a limiting parabolic profile. The interval is crossed quickly enough that silence ends most episodes. Accounting for possible returns, both the expected lifetime number of high-quality decisions and the probability of any such decision are of order $\delta^{3/2}$ from the specified upstream state. Theorems~\ref{thm:learning} and~\ref{thm:pulse} thus show why an equilibrium incentive can exist and still be rarely exercised near onset.

Preparation or initial service under a posted menu price provides one possible interpretation. Care devoted to a customer can change before a price is revised, and the customer may see an imperfect indication of that care before buying. Evidence on learning from other customers' choices and on infrequent restaurant price changes motivates these ingredients.\footnote{These studies examine observational learning and restaurant price adjustment: \citet{Cai2009}, \citet{MacDonaldAaronson2006}, and \citet{Fougere2010}.} The interpretation requires an operator who internalises continuation profit and a pricing authority restricted to the public record. We make neither restriction a claim about a named firm. Section~\ref{sec:discussion} explains the institution and the limits of the application.

Section~\ref{sec:literature} reviews related work. Sections~\ref{sec:model}--\ref{sec:core} develop the game, benchmark, and equilibrium construction. Section~\ref{sec:dynamics} states the learning, lifetime-quality, and local-adjustment results. Section~\ref{sec:discussion} discusses the economic interpretation, and Section~\ref{sec:conclusion} concludes. Appendices~A--D contain the proofs.

% END Introduction.tex

% BEGIN Literature.tex
\section{Related literature}\label{sec:literature}

Our paper relates to reputation for quality, pricing under social learning, and dynamic games with hidden actions. The specific question is how quality supplied at an inherited price survives equilibrium feedback and changes along the subsequent path of learning and price adjustment.

The role of prices in sustaining quality is well established. \citet{KleinLeffler1981} show how the prospect of losing future sales rents can deter contractual nonperformance. \citet{Shapiro1983} derives an equilibrium price--quality schedule in which quality premia also compensate sellers for investment in reputation. Our question concerns an inherited price that continues to support quality after it ceases to be optimal at a revision opportunity. Buyers learn from purchases and silence, and the incentive to supply quality changes as that learning proceeds. The contribution is the equilibrium construction and its adjustment properties, rather than the general proposition that a higher price can help sustain quality.

The informational role of prices is central to \citet{Bose.2006,Bose2008}. A monopolist sells to sequential buyers with private information, and price affects both current revenue and what later buyers can infer from purchase decisions. This creates a trade-off between current rent extraction and future learning. Our model builds on that mechanism. Here, however, quality is a private current choice of a strategic firm, and the public belief concerns a persistent type. The price affects the return to producing quality as well as the information generated by a purchase. Delayed revision makes the price determining this return an inherited part of the state.

\citet{Mahone.2024} connect pricing and observational learning to firm dynamics. In their benchmark, quality is exogenous and unknown to the firm as well as to consumers, while information cascades are an important source of dynamics. In our setting, the strategic type knows its type and chooses quality at individual customer opportunities. With full-support demand and strategic quality bounded away from one, a purchase remains informative at every interior belief. The adjustment we study can therefore occur while learning continues. Quality incentives change along a spell of silence even though the price is unchanged.

A close comparison is \citet{Pei2023}, who studies a strategic seller whose customers learn from predecessors' actions. We share the distinction between a commitment type and a strategic type choosing quality. His private-signal extension also allows customers to receive current private information when no past seller action is observed. This case is included in Section 5 of the \href{https://arxiv.org/abs/2006.08068}{2020 working-paper version}. Consequently, the absence of a public quality record is not by itself a distinction between the two models. Our focus is on delayed price choices and the resulting adjustment at fixed finite discounting. Public observation also differs: customer opportunities and individual refusals are hidden, so silence is informative without identifying a particular customer's action.

\citet{Knutsen.2025} endogenises price posting in a reputation model with limited records of quality. The allocation of pricing power changes equilibrium behaviour and can support efficient cycles. The pricing institution in our model instead assigns revision decisions to a desk that uses public information and maximises posterior expected firm profit. Quality is privately chosen by the strategic type. This separates the value governing price choice from the continuation value governing quality. It also rules out signaling through a privately informed price choice. Section~\ref{sec:model} states this institutional restriction.

Dynamic quality investment is studied by \citet{Board.2013}, and costly quality adjustment by \citet{Dilme2019}. These papers show how learning affects incentives to build and maintain quality. \citet{Kolb.2019} studies a buyer's adoption decision together with a seller's private decisions to upgrade persistent quality and exit. Depending on equilibrium expectations, investment may be motivated by a deteriorating reputation or by prospects of success. His analysis also features upgrade-cost cutoffs. Our quality choice is made separately at customer opportunities, without persistent upgrading or exit. We characterise its interaction with an inherited price and purchase-generated learning.

This distinction has implications for belief dynamics. In Kolb's model, anticipated hidden upgrading can itself raise the assessment of quality. In our model, the belief concerns the firm's persistent type. Along the branch we construct, silence lowers this belief, a purchase raises it, and a price reset leaves it unchanged. These three events consequently terminate a current quality episode in different ways. We connect these different routes of adjustment to the existence result and to the lifetime frequency of high-quality decisions.

Other work allows firms to influence the information used to assess them. \citet{Marinovic.2018} study certification, and \citet{Hauser.2024} studies costly promotion of quality. Reviews provide a further channel: \citet{Papanastasiou.2017} consider pricing with social learning and strategic consumers, while \citet{Chen.2025} study strategic pricing under rating systems. These environments make public information responsive to actions beyond the purchase itself.

The recent manuscript of Aleksenko and Kohlhepp (2026), \href{https://doi.org/10.2139/ssrn.4380756}{\emph{Reputational Underpricing}}, is particularly relevant to the pricing comparison. A privately informed firm sets prices continuously, and customers learn from value-based reviews and the current price, while past prices are unobserved. Price consequently serves both as a current signal and as an action affecting future reviews. They characterise conditions for underpricing and show that it need not occur even for a very patient firm. Quality in their model is exogenous and occasionally redrawn. Here quality is a hidden action, the persistent type is fixed, and the whole price history is public. A purchase supplies imperfect news; there is no public review or consumption outcome. Their signaling analysis illustrates why giving our producer private control of prices would change the game, rather than merely remove the revision friction.

The connection between quality and rigid prices goes back to \citet{Allen.1988}. His competitive industry has stochastic public demand and a moral-hazard constraint sustaining high quality. Buyers discover quality after purchase, and low quality destroys future sales in the equilibrium considered. Greater demand persistence can shift adjustment from prices towards quantities. Thus reputation can itself generate price rigidity. We take revision opportunities as given and examine their consequences for hidden quality under gradual Bayesian learning. There is no public revelation of an individual quality deviation, and no aggregate demand state whose persistence generates rigidity.\footnote{Allen also studies unobservable output, where the pattern of adjustment can differ. His results do not establish a universal positive relation between unobservable quality and price rigidity.} Experimental work likewise connects the two subjects: \citet{Renner.2004} study rigidity in customer markets, and \citet{Fehr2009} find that reputational incentives can improve contract enforcement while reducing rigidity.

\citet{Cripps.2004} establish the eventual limits of reputation under imperfect public monitoring. Our learning result supplies the endpoint of a particular adjustment process. The additional question is how much quality is supplied before that endpoint is reached. We link the cost gap to the local quality profile, the ways an episode ends, and the expected lifetime number of high-quality decisions, including possible returns to the active region. The lifetime bounds qualify the economic size of the existence result rather than merely establishing its eventual disappearance.

There is also a methodological connection to dynamic games with asymmetric information. \citet{Ouyang2025} study discrete-time stochastic games with hidden actions and derive Bayesian Nash equilibrium strategies based on compressed information. Our task is more specific: given the public-information pricing protocol, we construct stationary equilibria on the belief--price state space and characterise their paths. The local control of value feedback and the occupation bounds are used to establish this mechanism and its lifetime incidence; we do not offer a general existence theorem for dynamic games with hidden actions.

% END Literature.tex

% BEGIN Model.tex
\section{The model}\label{sec:model}
\subsection{The firm and the pricing decision}

Time is continuous, $t\geq0$. At date zero, the firm's type $\theta\in\{C,S\}$ is drawn, with prior $\mu_0=\Prb(\theta=C)\in(0,1)$. Type $C$ is committed to supplying high quality. Type $S$ chooses quality $x\in\{0,1\}$ at each customer opportunity. High quality, $x=1$, costs $c>0$ per opportunity; low quality is costless. The cost is incurred before the purchase decision and is paid even if the customer does not buy. The firm discounts future profit at rate $\rho>0$.

Customer opportunities arrive according to a Poisson process with rate $\nu>0$. Independent price-revision opportunities arrive at rate $\kappa>0$. The posted price belongs to a finite menu $P\subset(0,\bar p]$ and remains in force between revision opportunities. A purchase therefore does not automatically permit a price change.

At a revision opportunity, a pricing desk chooses a price from $P$ to maximise posterior expected firm profit. It observes the public history and can use independent randomisation. It does not observe the firm's type, its private quality decisions, unsuccessful customer opportunities, or private operating records. The strategic type controls quality but cannot override the desk's price choice.

The desk and the strategic producer evaluate firm profit using different information: the desk averages over the two possible types, while the producer knows its type. One interpretation is delegated pricing with a restricted reporting rule. An operating unit controls care and internalises the firm's continuation profit, while a price-setting authority uses the common transaction record. The model gives neither private reports nor operating accounts to that authority, and provides no private communication channel through which the producer can inform it. These restrictions define the game; they are not derived as an optimal organisational choice.

The desk is nevertheless forward-looking. It anticipates how a price affects demand, learning, and quality, and chooses the price giving the highest continuation profit at its information set. In particular, it does not maximise current revenue alone. Centralised pricing or delayed revision would not by themselves imply the information restriction.\footnote{An informed producer choosing prices would face a signaling problem; a platform maximising commission revenue would have a different objective. Neither is the pricing institution studied here.}

\subsection{Customers and purchase probabilities}

Write $\mu$ for the public probability of type $C$ and $a(\mu,p)$ for the assessed probability that type $S$ supplies high quality. The probability of current high quality is
\[
 \lambda=\mu+(1-\mu)a(\mu,p).
\]
This is a derived belief about the current product, whereas $\mu$ concerns the firm's persistent type. At an opportunity, the firm chooses quality before the customer observes a private signal $s\in\{H,L\}$ with precision $q\in(1/2,1)$:
\[
 \Prb(s=H\mid x=1)=\Prb(s=L\mid x=0)=q.
\]
The customer also observes an independent logistic taste shock with scale $\zeta>0$. High quality is worth $v>0$ and low quality is worth zero. The customer buys when expected quality value plus the taste shock is at least $p$.

Put $r_s=1-q$, $d=2q-1$, and $\Lambda(z)=e^z/(1+e^z)$. The signal posteriors and conditional purchase probabilities are
\begin{equation}\label{eq:buyer}
 \pi_H(\lambda)=\frac{q\lambda}{r_s+d\lambda},\qquad
 \pi_L(\lambda)=\frac{r_s\lambda}{q-d\lambda},\qquad
 b_s(\lambda,p)=\Lambda\!\left(\frac{v\pi_s(\lambda)-p}{\zeta}\right).
\end{equation}
Conditional on actual quality, the purchase probabilities are
\[
 A_H=qb_H+r_sb_L,\qquad A_L=r_sb_H+qb_L,\qquad D=A_H-A_L.
\]
Thus $D$ measures the increase in the probability of a sale from supplying high rather than low quality, with the customer's assessed rule held fixed. The strategic-type and public-mixture purchase probabilities are, respectively,
\[
 A_S=A_L+aD,\qquad \Abar=\mu A_H+(1-\mu)A_S=A_L+\lambda D.
\]
Unless otherwise stated, these objects are evaluated at $(\lambda,p)$.

A lower price increases demand, but need not increase its sensitivity to quality. Indeed,
\[
 \partial_pD=\frac d\zeta(b_H-b_L)(b_H+b_L-1).
\]
For fixed interior $\lambda$, the continuous-price extension of $D$ is maximised at $p=v(\pi_H+\pi_L)/2$. This follows because $b_H+b_L$ decreases strictly with price and equals one at that midpoint. The price that maximises current revenue need not coincide with the price that gives quality its largest effect on sales.

\subsection{Public information and reputation}

The public history records purchases, posted prices, all revision opportunities and their chosen prices, including decisions to keep the old price. Customer opportunities, individual non-purchases, signals, taste shocks, and consumption outcomes are not public. Elapsed time is observed. Since the two Poisson clocks are independent, simultaneous purchase and revision opportunities have probability zero.

For a stationary assessment, the public state is $(\mu,p)$. The memoryless revision clock makes price age unnecessary as an additional state. Reputation alone is not sufficient: at the same belief, two inherited prices can give different demand, learning, and quality incentives. Under types $C$ and $S$, purchases arrive at intensities $\nu A_H$ and $\nu A_S$. Bayes' rule therefore gives the silence drift and purchase update
\begin{equation}\label{eq:filter3}
 \beta=-\nu\mu(1-\mu)(1-a)D,\qquad
 J=\frac{\mu A_H}{\Abar}.
\end{equation}
The first formula follows by comparing the two probabilities of no purchase over a short interval; the second follows by conditioning on a purchase. Under the public mixture law,
\begin{equation}\label{eq:martingale}
 \beta+\nu\Abar(J-\mu)=0.
\end{equation}
When $a<1$ and belief is interior, silence reduces reputation and a purchase raises it. A reset leaves belief unchanged: at the desk's information set the price-choice distribution is the same under both types. This also specifies inference following a feasible off-path price choice by the desk. The public state consequently follows a piecewise deterministic process: belief drifts between public events, purchases change belief without changing the price, and revisions can change the price without changing belief. Section~\ref{sec:dynamics} uses this distinction to account for the three endings of a quality episode.

\subsection{Values and equilibrium}

Let $W$ and $V$ be continuation profit conditional on types $C$ and $S$. The desk evaluates $U=\mu W+(1-\mu)V$. Its reset kernel $r(p'\mid\mu,p)$ is supported on $\argmax_{p'}U(\mu,p')$; dependence on the old price is permitted for tie-breaking. Values are defined by expected discounted profit along the induced process. Their Bellman equations are understood along silence paths, allowing derivative jumps where the reset rule changes.

The normalisation
\[
 \Uti=U+\frac{\nu c\mu}{\rho}
\]
subtracts the committed type's unavoidable cost from price comparisons. The martingale identity \eqref{eq:martingale} implies
\begin{equation}\label{eq:UHJB3}
 \rho\Uti=\nu p\Abar-\nu c(1-\mu)a
 +\beta\Uti_\mu+\nu\Abar\{\Uti(J,p)-\Uti(\mu,p)\}
 +\kappa\{\max_{p'}\Uti(\mu,p')-\Uti(\mu,p)\}.
\end{equation}
For an exact maximising reset kernel, the strategic value satisfies
\begin{equation}\label{eq:VHJB3}
 \rho V=\nu(pA_S-ca)+\beta V_\mu
 +\nu A_S\{V(J,p)-V(\mu,p)\}
 +\kappa\left\{\sum_{p'}r(p'\mid\mu,p)V(\mu,p')-V(\mu,p)\right\}.
\end{equation}
The desk cannot choose when a revision opportunity arrives. Its maximum is consequently a continuation term multiplied by $\kappa$, rather than a condition allowing immediate intervention at every state.

At a private customer opportunity, the strategic firm's gain from high instead of low quality is
\begin{equation}\label{eq:obedience3}
 F=D\{p+V(J,p)-V(\mu,p)\}-c.
\end{equation}
The customer's behaviour, the public filter, and future strategies remain fixed when this deviation is evaluated. Only the likelihood of the current purchase changes. Recomputing beliefs as if the deviation were anticipated would give the firm a different commitment problem. Quality is optimal when $F\leq0$ at $a=0$, $F=0$ at $0<a<1$, and $F\geq0$ at $a=1$.

A stationary equilibrium consists of a quality policy, a type-independent reset kernel, continuation values, and Bayesian beliefs satisfying these conditions at every state and current price. We require a well-defined public process; the construction below has Lipschitz quality policies and nonexplosive paths. The committed type follows its prescribed action. Short-lived customers optimise according to \eqref{eq:buyer}; the desk chooses an exact value maximiser; and the strategic quality policy is optimal given the assessment.\footnote{Values can equivalently be defined through the discounted payoff integrals. This avoids imposing differentiability at price switches. The appendix verifies the Bellman inequalities against alternative private quality strategies, including history-dependent deviations.}

% END Model.tex

% BEGIN Benchmark.tex
\section{The zero-quality benchmark}\label{sec:benchmark}

Consider first an assessment in which the strategic type always supplies low quality. Then $a=0$, $\lambda=\mu$, and we use a superscript zero for the resulting purchase probabilities, filter, and values. Equation~\eqref{eq:UHJB3} no longer contains $c$. The normalised desk value and its optimal reset correspondence are therefore independent of the production cost. Conditional on a chosen exact desk selector $r$, the strategic value $V^{0,r}$ is also independent of $c$.

For completeness, the desk's next-event operator is a contraction on bounded continuous functions, with modulus at most $(\nu+\kappa)/(\rho+\nu+\kappa)<1$. The silence flow and purchase probabilities are smooth, and the price menu is finite. Its fixed point defines the normalised desk value; a measurable exact maximiser exists. Fixing such a selector gives a similar contraction for the strategic payoff. These observations make the benchmark a well-defined problem rather than an assumption about unknown equilibrium values.

Define the gross gain from a high-quality deviation by
\[
 \Gamma^{0,r}(\mu,p)=D^0(\mu,p)
 \{p+V^{0,r}(J^0(\mu,p),p)-V^{0,r}(\mu,p)\},
 \qquad c^*(r)=\sup_{\mu,p}\Gamma^{0,r}(\mu,p).
\]
Substitution in \eqref{eq:obedience3} gives the exact threshold: zero strategic quality is sequentially rational at every state if and only if $c\geq c^*(r)$. At $c>c^*(r)$ the obedience margin is uniformly strict. This establishes existence of a zero-quality equilibrium, not that every equilibrium has zero quality at these costs. If ties affect strategic continuation values, the threshold can depend on the selector.\footnote{In the primitive family used below there is one isolated interior price tie. The belief process spends zero event-time measure at that point, so its tie choice does not affect values. This is sufficient for selector independence; a globally unique price choice is not required.}

The threshold ranges over all current prices. To understand why this matters, suppose tie choices are immaterial and write $\Gamma^0$ and $c^*$. Compare it with
\begin{equation}\label{eq:reset3}
 c^*_{\rm reset}=\max\{\Gamma^0(\mu,p):p\in\argmax_{p'}\Uti^0(\mu,p')\}.
\end{equation}
The maximum is well-defined for the continuous benchmark constructed below and includes every tied maximiser. If $c^*_{\rm reset}<c<c^*$, high quality is unprofitable at every reset-optimal price under the zero-quality assessment, but profitable at some inherited price. Hence that assessment cannot be an equilibrium throughout the state space.

This comparison locates the incentive that can initiate positive quality. It does not yet establish that the incentive survives when customers anticipate quality and the desk reoptimises. It also keeps the continuation values and reset technology fixed. A fully flexible-price economy would require its own equilibrium analysis. The next section provides a primitive construction in which the inherited-price incentive survives the full equilibrium feedback.

% END Benchmark.tex

% BEGIN Results.tex
\section{An equilibrium with quality at an inherited price}\label{sec:core}

We first construct the equilibrium whose adjustment is studied in Section~\ref{sec:dynamics}. The high price must be an earlier optimal choice, quality must require a continuation reward, and the desk must prefer the low price in the active region. These conditions exclude an arbitrary inherited price and a purely static return to quality. Theorem~\ref{thm:primitive} and Corollary~\ref{cor:essential} establish them for the following analytically tractable two-price family; Proposition~\ref{prop:robustness} extends the result to an open set of primitives:
\begin{equation}\label{eq:primitives3}
 \zeta=1,\qquad v=6,\qquad q=\frac{11}{20},\qquad
 P=\left\{p_L=\frac52,\ p_H=3\right\},\qquad \nu=\kappa=1.
\end{equation}
Set
\[
 D_* =\frac1{10}\tanh\!\left(\frac3{20}\right),\qquad c_{\rm dir}=3D_*.
\]
The quantity $c_{\rm dir}$ is the largest immediate quality return $pD(\lambda,p)$ over all $\lambda\in[0,1]$ and both prices.

\begin{theorem}\label{thm:primitive}
For the primitives \eqref{eq:primitives3}, there is a finite $\bar\rho$ such that the following statements hold for every fixed $\rho>\bar\rho$.
\begin{enumerate}[label=(\roman*)]
\item The zero-quality desk has one simple interior switch, from $p_L$ to $p_H$ as belief rises. The gain $\Gamma^0$ has a unique maximum at $(\mu_*,p_H)$, with $\mu_*\to1/2$ as $\rho\to\infty$, and
\[
 c^*>\max\{c^*_{\rm reset},c_{\rm dir}\}.
\]
\item There is $\bar\delta(\rho)>0$ such that, for every $c=c^*-\delta$, $0<\delta<\bar\delta(\rho)$, the original unstopped model has a stationary equilibrium $(a_\delta,r_\delta,U_\delta,V_\delta)$ with exact Bayesian updating and exact quality and desk obedience. Its strategic quality is positive at $(\mu_*,p_H)$, zero at $p_L$, and has support contained in an interior interval of width $O(\sqrt\delta)$, with
\begin{equation}\label{eq:branch3}
 \|a_\delta\|_\infty=O(\delta),\quad
 \|a_\delta\|_{1,P}=O(\delta^{3/2}),\quad
 \|\Uti_\delta-\Uti^0\|_\infty+\|V_\delta-V^0\|_\infty=O(\delta^{3/2}).
\end{equation}
Here $\|f\|_{1,P}=\max_{p\in P}\int_0^1|f(\mu,p)|\,d\mu$.
\item Throughout the effort region the desk uniquely resets to $p_L$. The active inherited price $p_H$ is uniquely reset-optimal at $\mu_U=3/4$, and the effort region is reached with positive probability from $(\mu_U,p_H)$.
\end{enumerate}
All constants in the $\delta$ estimates hold $\rho$ and the other primitives fixed. No uniqueness of the positive equilibrium is asserted.
\end{theorem}

\begin{corollary}\label{cor:essential}
After further restricting
\[
 0<\delta<\min\{\bar\delta(\rho),\ c^*-c_{\rm dir},\ c^*-c^*_{\rm reset}\},
\]
the equilibria in Theorem~\ref{thm:primitive} satisfy
\[
 pD(\lambda,p)-c<0\quad\hbox{for every }(\lambda,p)\in[0,1]\times P.
\]
Thus their positive strategic quality requires the continuation reward
$D\{V_\delta(J,p)-V_\delta(\mu,p)\}$. An immediate reset in the effort region selects a price at which equilibrium quality is zero.
\end{corollary}

The immediate return helps determine where quality is attractive, but cannot cover its cost at any state in this interval. A reputational continuation reward is therefore essential. The reset comparison holds within the constructed equilibrium; Section~\ref{sec:discussion} explains its relation to a change in the pricing institution.

The proof proceeds from the benchmark geometry to the equilibrium feedback. The zero-quality desk has a simple price switch, while the maximum quality gain occurs strictly below it at the high price. The continuation reward raises this gain above the largest immediate return. Lowering cost by $\delta$ creates an active region of width $O(\sqrt\delta)$ with quality probability $O(\delta)$. The resulting $O(\delta^{3/2})$ mass controls value changes and displacement of the desk's switch. This is small enough to preserve the strict price ranking in the active region. A fixed point then imposes exact quality and desk obedience jointly. Appendix~\ref{sec:baselineproof} proves the primitive inequalities, and Appendix~\ref{sec:existence} gives the construction. The same height and width will determine the local dynamics in Theorem~\ref{thm:pulse}.

The order of quantifiers matters. The discount rate is fixed at a sufficiently large finite value before the cost shortfall is made small. In fact, Proposition~\ref{prop:benchmark} shows that the gap $c^*-c_{\rm dir}$ is positive but of order $1/\rho$. The cost interval requiring reputation therefore shrinks as the firm becomes increasingly impatient. The theorem does not give estimates uniform in a simultaneous limit of the discount rate and the cost gap.

% BEGIN Robustness_Result.tex
\subsection{Robustness to primitive changes}

The particular constants in \eqref{eq:primitives3} provide an analytically verified economy. They do not restrict the result to that economy. To state the extension, identify prices by their labels $L,H$ and write
\[
 \xi=(\zeta,v,q,p_L,p_H,\nu,\kappa,\rho),\qquad
 \xi_0=(1,6,11/20,5/2,3,1,1,\rho_0).
\]
The admissible parameter space has positive taste scale, valuation, arrival rates and discount rate, $1/2<q<1$, and $0<p_L<p_H$. For each parameter vector, define $c^*(\xi)$ from its own zero-quality benchmark and put
\[
 c_{\rm dir}(\xi)=\max_{\lambda\in[0,1],\ i\in\{L,H\}}
 p_iD_\xi(\lambda,p_i).
\]
The reset-restricted threshold $c^*_{\rm reset}(\xi)$ likewise uses that economy's benchmark desk and includes tied reset choices.

\begin{proposition}\label{prop:robustness}
Fix any sufficiently large finite $\rho_0$ for which Theorem~\ref{thm:primitive} applies. There are an open neighbourhood $\mathcal N$ of $\xi_0$, with compact closure in the admissible parameter space, and $\delta_0>0$ such that the following statements hold for every $\xi\in\overline{\mathcal N}$.
\begin{enumerate}[label=(\roman*)]
\item The zero-quality desk has exactly one simple interior switch. The benchmark gain has a unique interior maximum $(\mu_*(\xi),p_H)$ with strictly negative belief curvature and strictly negative trial-quality slope. Moreover,
\[
 c^*(\xi)>\max\{c^*_{\rm reset}(\xi),c_{\rm dir}(\xi)\}.
\]
The switch, maximum, thresholds, and local curvature and slope depend continuously on $\xi$.
\item For every $0<\delta<\delta_0$, at cost $c=c^*(\xi)-\delta$ the original model has a stationary equilibrium with positive quality at $(\mu_*(\xi),p_H)$ and zero quality at $p_L$. The support at $p_H$ lies where the desk uniquely resets to $p_L$, and it is reached with positive probability from the optimally posted state $(3/4,p_H)$. Every immediate quality return is strictly below cost. The height, support-width, effort-mass and value bounds in \eqref{eq:branch3} hold with constants common to $\overline{\mathcal N}$.
\item For the equilibria constructed in (ii), Theorems~\ref{thm:learning} and~\ref{thm:pulse} hold with the primitives and zero-quality objects of that economy. The local remainders and lifetime upper bounds are uniform on $\overline{\mathcal N}$, and the lifetime $\Theta(\delta^{3/2})$ bounds from $(3/4,p_H)$ have common positive lower and upper constants. For a lifetime upper bound from another state, the initial interior belief and the price label are held fixed.
\end{enumerate}
\end{proposition}

Appendix~\ref{sec:robustproof} proves the proposition by controlling the moving benchmark switch and then constructing an equilibrium separately for each nearby economy. No continuous or unique selection of positive-quality equilibria is assumed. The cost gap is always measured from the threshold of the perturbed economy, $c^*(\xi)$, rather than from the original threshold. The neighbourhood lies within the two-price, logistic-demand model with its public-information pricing protocol. It is local to the fixed finite $\rho_0$; no uniform neighbourhood as $\rho_0\to\infty$ is asserted.

% END Robustness_Result.tex

% END Results.tex

% BEGIN Dynamics.tex
\section{Equilibrium adjustment and lifetime quality}\label{sec:dynamics}

We now follow the equilibria constructed in Section~\ref{sec:core}. There are two distinct questions: whether quality eventually disappears, and how much is supplied before it does. Theorem~\ref{thm:learning} answers them over the whole equilibrium path. Theorem~\ref{thm:pulse} then describes a single traversal of the active region and explains the order of the lifetime response. Both results hold for the same local family; Proposition~\ref{prop:robustness} extends them to nearby economies, measuring each cost gap from that economy's own threshold. Proofs are in Appendix~\ref{sec:dynamicproofs}.

\subsection{Learning, cessation, and lifetime incidence}

\begin{theorem}\label{thm:learning}
In any well-defined Bayesian assessment with a finite price menu, full-support logistic demand, $q>1/2$, and $a\leq\bar a<1$, the public posterior satisfies
\[
 \mu_t\longrightarrow\ind_{\{\theta=C\}}\quad\hbox{almost surely}.
\]
For the equilibria in Theorem~\ref{thm:primitive}, strategic quality becomes identically zero after a finite random time. The posted price eventually becomes permanently $p_L$ for the strategic type and $p_H$ for the committed type, almost surely.

If $N_H$ counts the strategic firm's high-quality decisions over its entire lifetime, then, for any fixed initial interior state,
\[
 \E_S N_H=O(\delta^{3/2}),\qquad
 \Prb_S(N_H\geq1)=O(\delta^{3/2}).
\]
From the optimally posted upstream state $(3/4,p_H)$ both quantities are $\Theta(\delta^{3/2})$.
\end{theorem}

The learning assertion uses persistent informativeness of purchases: on every compact interior belief interval, the purchase intensities of the two types remain separated when strategic quality is bounded away from one. The quality-support and endpoint-price properties of the constructed branch then give cessation and permanent price selection. The committed type continues to supply quality. Almost-sure cessation does not imply a finite expected last-effort time, and the theorem gives no uniform exponential learning rate.

The lifetime bounds require more than eventual learning or a calculation for one visit. A purchase can return belief to the active region, so all subsequent visits must be counted. The proof controls their total expected occupation and combines this with the $O(\delta)$ quality probability. From the upstream state there is a positive limiting chance of reaching the region, which gives the matching lower bounds. Near the threshold, a strategic firm is thus unlikely ever to incur the quality cost, despite the positive probability of encountering a state at which quality is optimal with positive probability.

\subsection{Local quality episodes and their endings}

Consider entry into the smallest interval containing the high-price quality support. We call its first traversal an episode, ending at the first purchase, the first reset, or exit through continued silence. This definition does not require connected support or a realised high-quality decision during the episode. A reset ends the current episode but need not preclude later episodes; only the learning result gives eventual permanent cessation.

To describe the local profile, hold a continuation value $V$ fixed and let $t$ be a trial probability of strategic quality at price $p_H$. Write $F_V(\mu,t)$ for the gain in \eqref{eq:obedience3} evaluated at $\lambda=\mu+(1-\mu)t$ and the corresponding purchase update. Thus the trial derivative below changes the assessment used in demand and filtering while holding $V$ fixed. It is distinct from evaluating a unilateral hidden-action deviation, for which the assessment remains fixed.

\begin{theorem}\label{thm:pulse}
Fix $\rho>\bar\rho$. Define
\[
 K=-\Gamma^0_{\mu\mu}(\mu_*,p_H)>0,\qquad
 m=-\partial_tF_{V^0}(\mu_*,0)>0,\qquad Y=\sqrt{2/K},
\]
where $t$ is a trial assessment holding $V^0$ fixed. Let
\[
 b_*=-\beta^0(\mu_*,p_H)>0,\qquad h_S^*=\nu A_L^0(\mu_*,p_H).
\]
For every equilibrium furnished by Theorem~\ref{thm:primitive}, uniformly on bounded sets of $y$,
\begin{equation}\label{eq:pulse3}
 \frac{a_\delta(\mu_*+\sqrt\delta\,y,p_H)}{\delta}
 \longrightarrow\frac1m\pos{1-\frac K2y^2}.
\end{equation}
Let $[\ell_\delta,u_\delta]$ be the smallest closed interval containing the support of $a_\delta(\cdot,p_H)$. Its endpoints satisfy
\[
 \ell_\delta=\mu_*-Y\sqrt\delta+o(\sqrt\delta),\qquad
 u_\delta=\mu_*+Y\sqrt\delta+o(\sqrt\delta).
\]
We do not require the support itself to be connected. Starting at $(u_\delta,p_H)$, define the first traversal to end at the first purchase, the first reset, or the first time belief reaches $\ell_\delta$, whichever occurs first. The no-purchase, no-reset crossing time of this interval is
\[
 T_\delta=\frac{2Y}{b_*}\sqrt\delta+o(\sqrt\delta).
\]
A purchase jumps above the active region; a reset selects the inactive price. Under the strategic-type law, the first traversal ends by reset, purchase, or silent exit with respective probabilities
\begin{align*}
 P_{\rm reset}&=\frac{2\kappa Y}{b_*}\sqrt\delta+o(\sqrt\delta),\\
 P_{\rm purchase}&=\frac{2h_S^*Y}{b_*}\sqrt\delta+o(\sqrt\delta),\\
 P_{\rm silent}&=1-\frac{2(\kappa+h_S^*)Y}{b_*}\sqrt\delta+o(\sqrt\delta).
\end{align*}
Expected high-quality decisions during that traversal, including a possible terminating purchase, equal
\begin{equation}\label{eq:pulsecount3}
 \frac{4\nu Y}{3mb_*}\delta^{3/2}+o(\delta^{3/2}).
\end{equation}
\end{theorem}

The parabolic profile follows from the quadratic maximum of the benchmark quality gain and its strictly negative response to an increase in assessed quality. The gain's curvature determines the width, while the trial-quality slope determines the height. Silence then carries belief through this interval at a speed bounded away from zero. Multiplying the $O(\delta)$ quality probability by the $O(\sqrt\delta)$ crossing time explains the $\delta^{3/2}$ traversal count.

The three endings separate changes in information from changes in the price. A purchase raises reputation beyond the narrow region while leaving the price fixed. A revision selects the low price and removes the incentive without changing reputation. Silence carries reputation below the region, again at the old price. Its dominance near onset reflects the short crossing time relative to the two public-event clocks, not the absence of informative purchases. The fixed reset intensity and the limiting strategic purchase intensity determine the first-order probabilities of the other endings.

This local calculation and the lifetime result are complementary. The first describes the shape of a quality episode and its termination; the second establishes that repeated visits do not change the order of total quality provision. The limits hold at fixed finite discounting and fixed reset intensity. They provide neither a global comparative static in the revision rate nor a welfare comparison across pricing institutions.

% END Dynamics.tex

% BEGIN Discussion.tex
\section{Economic interpretation}\label{sec:discussion}
\subsection{The pricing institution and a service-market interpretation}

The economic comparison requires two restrictions together. Quality is privately adjustable at a customer opportunity, while the operating unit has no discretion over the posted price. In addition, the price-setting authority evaluates continuation profit from the public record, not from the producer's private type or operating information. The desk's posterior objective and the producer's type-conditional objective can therefore rank their available choices differently. The result concerns behaviour within this delegated arrangement, not a firm's choice to adopt it.

The food-service illustration concerns care devoted to an offer or to the initial customer encounter before the decision to buy. Its timing matters: the quality cost is paid even without a sale. Preparation or attention directed at a particular customer is therefore a closer interpretation than quality supplied only after payment. A lasting investment or a batch serving many customers would require an additional state for quality or inventory. The operator must also internalise the firm's continuation profit; a fixed-wage employee without such an incentive would face a different problem.

Several empirical findings motivate parts of this setting. \citet{Cai2009} show that information about the popularity of dishes affects diners' choices, separately from drawing attention to the dishes. \citet{MacDonaldAaronson2006} and \citet{Fougere2010} document infrequent restaurant price changes and examine adjustment following minimum-wage increases.\footnote{Nearly uniform prices across outlets in \citet{DellaVigna.2019} also motivate limits on local price discretion. Spatial uniformity is distinct from delayed adjustment over time, and a common price across heterogeneous outlets need not solve the single-outlet problem considered here.} These findings support learning from choices and delayed price adjustment as relevant ingredients. They do not establish the exact transaction record, the independent revision clock, or the information available to the desk.

This is an interpretation of the timing, not an empirical validation of the full institution. We do not identify a named chain that restricts pricing to public information; \citet{Zbaracki2004} in fact document organisational adjustment costs alongside the use of internal information. Visible patronage or a popularity ranking also differs from an exact transaction-time record. A substantive application would need to establish who controls prices, whose information enters that decision, and whether the operator bears the quality cost and receives the continuation return. Reviews, inspections, queues, and repeat customers would require accounting for additional information.

\subsection{The meaning of the reset comparison}

Corollary~\ref{cor:essential} compares choices within one equilibrium. At a state with positive strategic quality, an immediate revision selects the low price, where equilibrium quality is zero. Belief is unchanged at the revision because the desk uses public information. The comparison therefore isolates the role of the current price while preserving the game's information structure and continuation strategies.

Eliminating the revision friction would change future prices, incentives, and learning throughout the state space. Such an economy requires its own equilibrium construction. Our result consequently does not establish that greater price rigidity raises aggregate quality or welfare. It shows that a price inherited from an earlier optimal choice can sustain a temporary quality incentive, and describes how that incentive disappears under the given revision technology.

\subsection{Quality provision and its measurement}

The existence and rarity results should be read together. Quality occurs on an open set of primitives and requires a continuation reward, yet the probability that the strategic firm ever supplies it tends to zero as cost approaches the threshold. The construction establishes an incentive that survives equilibrium feedback without claiming a large quantitative response.

The lifetime count concerns quality decisions at customer opportunities, including those without a sale. Measuring only completed transactions would select on quality. At a fixed state of the strategic firm, the probability that a purchased unit is high quality is
\[
 \Prb(x=1\mid\text{purchase},S,\mu,p)
 =\frac{a A_H}{A_L+aD}>a\qquad\text{if }0<a<1.
\]
The inequality follows from $D>0$ at an interior assessment. An empirical investigation would therefore need information on opportunities and unsuccessful encounters, as well as prices and sales. Evidence based on sales alone cannot directly measure the lifetime count in Theorem~\ref{thm:learning}.

% END Discussion.tex

% BEGIN Conclusion.tex
\section{Conclusion}\label{sec:conclusion}

A price inherited from an earlier reputation can sustain hidden quality after the pricing desk would choose to lower it. We construct such an equilibrium on an open set of primitives, with quality costs above every immediate sales return. The distinction between the inherited price and the reset choice is therefore supported by a continuation reward, not simply by a high current margin. It survives the response of buyers, beliefs, values, and subsequent price choices to anticipated quality.

The same equilibrium construction permits a precise account of adjustment. Near the threshold, the quality probability has a parabolic local profile. Purchases end episodes by raising reputation, revisions by changing the price, and silence by lowering reputation; the last route dominates near onset. Accounting for possible returns gives lifetime quality of order $\delta^{3/2}$ from the specified upstream state. Eventually the persistent type is learned, strategic quality ceases, and prices settle at type-specific levels. Existence, local adjustment, and lifetime incidence thus describe different aspects of one temporary incentive.

The model takes the pricing protocol and revision opportunities as given. Allowing private information to enter price choices would introduce signaling, while endogenous revision opportunities would let the firm choose its adjustment technology. These extensions would require a new equilibrium analysis to compare price flexibility, quality, and welfare.

% END Conclusion.tex

\appendix

% BEGIN Proofs.tex
% Complete proof draft for the analytical core.
\section{Primitive geometry and finite-discount values}\label{sec:baselineproof}

Throughout this section the assessment is zero and the primitives are \eqref{eq:primitives3}. Write $R_i(\mu)=p_i\Abar^0(\mu,p_i)$ and $S_i(\mu)=p_i A_L^0(\mu,p_i)$, for $i=L,H$. Constants in large-$\rho$ bounds depend only on these fixed primitives and specified compact intervals. They do not depend on a selected zero-quality tie action.

\begin{lemma}\label{lem:geometry}
The following primitive properties hold.
\begin{enumerate}[label=(\roman*)]
\item $D(\lambda,p)\leq D_*$ for every $\lambda\in[0,1]$ and price $p$. Equality holds only at $(\lambda,p)=(1/2,3)$. The maximum of $3D(\cdot,3)$ is strictly quadratic, and $D(\cdot,3)$ decreases strictly to either side of its midpoint maximum.
\item $R_H-R_L$ has exactly one zero $\tau_0\in(0,1)$, and its derivative there is strictly positive.
\item On $I=[12/25,13/25]$, the low price has a uniform revenue advantage
\begin{equation}\label{eq:I-margin}
 R_L-R_H>\epsilon_I:=\frac{2181127}{533688000}>0.
\end{equation}
At $\mu_U=3/4$ the high price has a strict revenue advantage.
\item If $G(\mu)=\mu\partial_\mu A_L(\mu,3)$, then
\begin{equation}\label{eq:weightedprimitive}
 G'(\mu)>\frac{5309}{100000}>0\qquad (\mu\in I).
\end{equation}
\end{enumerate}
\end{lemma}
\begin{proof}
The posterior gap is
\[
 \pi_H-\pi_L=\frac{d\lambda(1-\lambda)}{qr_s+d^2\lambda(1-\lambda)}\leq d,
\]
with equality only at $\lambda=1/2$. The logistic probability of an interval of length $x$ is at most $\tanh(x/4)$, with equality only when the interval is centered at zero. This proves (i)'s bound and equality statement. At price $3=v/2$, moving away from $1/2$ shortens the interval and moves its center away from zero, strictly reducing the difference. With $y=\lambda-1/2$,
\[
 \pi_H-\pi_L=d\frac{1-4y^2}{1-4d^2y^2}
 =d-4d(1-d^2)y^2+O(y^4).
\]
Smoothness, symmetry, and the resulting strict quadratic loss prove the curvature assertion. For example, putting $s=\Lambda'(3/10)$ and $T=\tanh(3/20)$ gives the explicit check
\[
 D_{\mu\mu}(1/2,3)
 =-72d(1-d^2)^2sT-48d^2(1-d^2)s<0.
\]

For (ii), compare revenues conditional on a signal posterior $\pi$. Put $t=e^{1/2}$ and $y=e^{6\pi-3}$. The high-price advantage is
\[
 f(y)=\frac{y(A+By)}{(1+y)(1+ty)},\quad A=3-\frac52t<0,\quad B=\frac12t>0.
\]
Since $3/2<t<2$, its unique sign change is at $y_0=-A/B=5-6/t\in(1,2)$, or $\pi_0=(3+\log y_0)/6\in(0,1)$. Moreover
\[
 f'(y)=\frac{A+2By+\{B(1+t)-At\}y^2}{(1+y)^2(1+ty)^2}>0
 \quad\hbox{when }y\geq y_0/2.
\]
Only beliefs with $\pi_L(\mu)\leq\pi_0\leq\pi_H(\mu)$ can be zeros of aggregate revenue difference. On this interval,
$e^{6\pi_L-3}\geq y_0e^{-3/5}>y_0/2$; the last inequality follows from $\log2>2/3>3/5$. Both conditional advantages therefore increase strictly with belief. Their weights are $w_H=r_s+d\mu$ and $w_L=q-d\mu$, so the derivative of their weighted sum is
\[
 d\{f(y_H)-f(y_L)\}+w_H\frac{d f(y_H)}{d\mu}
 +w_L\frac{d f(y_L)}{d\mu}>0.
\]
Below this interval both advantages are negative; above it both are positive. This proves a single simple crossing.

For (iii), $R_H(1/2)=3/2$. Taylor lower bounds on the exponential give
\[
 e^{4/5}>\frac{827}{375},\qquad e^{1/5}>\frac{61}{50},\qquad
 R_L(1/2)-R_H(1/2)>
 \frac54\left(\frac{827}{1202}+\frac{61}{111}\right)-\frac32
 =\frac{25063}{533688}.
\]
Globally, $|\Abar_\mu|\leq29/15$ and
\[
 |\partial_p\Abar_\mu|\leq\frac d4+\frac{v(q/r_s)}8=\frac{113}{120}.
\]
Here we used $\Lambda'\leq1/4$, $|\Lambda''|\leq1/8$, and $\pi_s'\leq q/r_s$. Decomposing the price difference yields
\[
 |(R_L-R_H)'|\leq\frac12\frac{29}{15}+\frac52\frac12\frac{113}{120}
 =\frac{343}{160}.
\]
Since $I$ has half-width $1/50$, its revenue margin exceeds
$25063/533688-343/8000=\epsilon_I$.
At $\mu=3/4$, $\pi_H=11/14$ and $\pi_L=27/38$. For both signals $y=e^{6\pi_s-3}>2>y_0$, so the high-price advantage is positive.

For (iv), on $I$ one has $49/50<\pi_s'<1$, $|6\pi_s-3|\leq105/251<1/2$, and $|\pi_H''|<41/100$. Thus $\Lambda'(6\pi_s-3)>2/9$ and
\[
 A_L'>6\frac{49}{50}\frac29>\frac54.
\]
Also $6\pi_L-3<0$ and $\pi_L''>0$, so both low-signal terms in $A_L''$ are positive. Bounding only the negative high-signal terms gives
\[
 A_L''\geq-\frac9{20}\left(\frac{36}{8}+\frac64\frac{41}{100}\right)
 =-\frac{9207}{4000}.
\]
Consequently $G'=A_L'+\mu A_L''>5/4-(13/25)(9207/4000)=5309/100000$.
\end{proof}

\begin{lemma}\label{lem:regularity}
For sufficiently large $\rho$, the zero-quality desk has exactly one simple switch $\tau_\rho$, with $\tau_\rho=\tau_0+O(\rho^{-1})$. Its normalized value is globally Lipschitz with norm $O(\rho^{-1})$, is $C^1$ in the interior, and satisfies, on compact interior intervals,
\begin{equation}\label{eq:U-C1}
 \rho\Uti_i^0=R_i+O(\rho^{-1}),\qquad
 \rho(\Uti_i^0)'=R_i'+O(\rho^{-1}).
\end{equation}
The strategic value is globally Lipschitz with norm $O(\rho^{-1})$, continuous at the endpoints, and $C^1$ away from $\tau_\rho$. Uniformly on compact intervals separated from the switch,
\begin{equation}\label{eq:V-C1}
 \rho V_i^0=S_i+O(\rho^{-1}),\qquad
 \rho(V_i^0)'=S_i'+O(\rho^{-1}).
\end{equation}
Where both $\mu$ and $J_i^0(\mu)$ stay away from the switch, $V_i^0$ is $C^2$ and $(V_i^0)''=O(\rho^{-1})$ on compact intervals.
\end{lemma}
\begin{proof}
We give the regularity argument because convergence of value levels alone does not prove a simple switch or gain curvature. All zero-quality drifts, hazards, rewards, and purchase maps are smooth on $[0,1]$; the purchase denominator is bounded below by $\min_p\Lambda(-p)>0$. Let $\phi_i(t,\mu)$ be the silence flow. For a primitive constant $B$,
\[
 |\partial_\mu\phi_i(t,\mu)|\leq e^{Bt},\qquad
 |\partial_\mu\phi_i(t,\mu)-1|\leq Bt e^{Bt}.
\]
The next-event Bellman operator for the aggregate value is
\begin{align}\label{eq:firstU}
 (T^Uw)_i(\mu)=\int_0^\infty e^{-\rho t}H_i^U(t,\mu)
 \big[&R_i(\phi_i)+\Abar_i(\phi_i)w_i(J_i(\phi_i))\notag\\
 &+\max_jw_j(\phi_i)\big]dt,
\end{align}
where $H_i^U=\exp[-\int_0^t(1+\Abar_i(\phi_i(s,\mu)))ds]$ and the suppressed flow argument is $(t,\mu)$. Its continuation mass is at most $2/(\rho+2)$, so it is a sup-norm contraction on continuous functions. Its fixed point is the desk's optimal value, by iteration and the finite-menu maximizing rule.

For bounded Lipschitz $w$, direct differentiation almost everywhere, using $\Lip(\max_jw_j)\leq\max_j\Lip(w_j)$, gives constants $C$ independent of $\rho$ such that
\begin{equation}\label{eq:lipU}
 \Lip(T^Uw)\leq\frac C\rho+\frac C{\rho-B}\{\|w\|_\infty+\Lip(w)\}
 \qquad (\rho>B+1).
\end{equation}
Indeed, derivatives of the survival factor are bounded by $Ct e^{Bt}$, and derivatives of the flow and of the smooth purchase map by $Ce^{Bt}$. Integrating these bounds proves \eqref{eq:lipU}, after enlarging $B,C$ if necessary. Picard iteration from zero has sup norm at most $3/\rho$. Taking $\rho$ so that the Lipschitz coefficient is below one proves $\Lip(\Uti^0)=O(\rho^{-1})$.

The reward-only integral in \eqref{eq:firstU} has derivative $R_i'(\mu)/\rho+O(\rho^{-2})$: subtract its value at $t=0$ inside the integral and use the smoothness and flow estimates above. The continuation part has derivative $O(\rho^{-2})$, since both the norm and the Lipschitz modulus of the fixed point are $O(\rho^{-1})$. Thus the derivative expansion first holds almost everywhere. In the interior, the Bellman equation expresses $\beta_i(\Uti_i^0)'$ as a continuous function; $\beta_i<0$ there. It follows that $\Uti_i^0$ is $C^1$ and the expansion holds everywhere. The analogous level estimate is uniform on $[0,1]$.

Uniform convergence of $\rho(\Uti_H^0-\Uti_L^0)$ confines its zeros to any fixed small neighborhood of $\tau_0$. On such a neighborhood the derivative of $R_H-R_L$ has a positive minimum. Equation \eqref{eq:U-C1} makes the discounted desk difference strictly increasing there. Hence there is precisely one simple switch and its displacement is $O(\rho^{-1})$. It stays in a fixed interior interval, so the silence speeds there have a common positive lower bound.

For the strategic operator replace $R_i,\Abar_i,\max_jw_j$ in \eqref{eq:firstU} by $S_i,A_{L,i},w_{R_\rho(\phi_i)}$, where $R_\rho=L$ below $\tau_\rho$ and $H$ above it. The last term is piecewise Lipschitz and can jump by $w_H(\tau_\rho)-w_L(\tau_\rho)$. If a silence path crosses the switch at time $T_i(\mu)$, differentiation of the two pieces of the integral contributes a boundary term bounded by
\begin{equation}\label{eq:switchterm}
 C\|w\|_\infty e^{-\rho T_i(\mu)}|T_i'(\mu)|
 \leq C\|w\|_\infty e^{-(\rho-B)T_i(\mu)}.
\end{equation}
For the inequality, differentiate $\phi_i(T_i(\mu),\mu)=\tau_\rho$ and use the nonzero speed at the switch. There is no such term for a path starting below the switch. The hitting-time formula also controls starting beliefs approaching an endpoint.

The resulting global Lipschitz estimate is
\[
 \Lip(T^Vw)\leq\frac C\rho+C\|w\|_\infty
 +\frac C{\rho-B}\Lip(w).
\]
Picard iteration, again bounded by $3/\rho$, gives $\Lip(V^0)=O(\rho^{-1})$. The operator preserves continuity: integration across a switch moves an integration endpoint, and no jump in the value is introduced. Its unique fixed point is continuous on the closed state space. At a compact positive distance above the switch, the crossing time has a positive lower bound; hence \eqref{eq:switchterm} is $O(\rho^{-2})$ or smaller. Below it the term vanishes. The same reward/continuation decomposition now gives \eqref{eq:V-C1}. The strategic Bellman equation yields $C^1$ regularity wherever the reset rule is constant.

Finally, at a point where $\mu$ and $J_i(\mu)$ avoid the switch, differentiate that Bellman equation. With $h_i=A_{L,i}$ and $R=R_\rho(\mu)$ it gives
\[
 \beta_i(V_i^0)''=(\rho+h_i+1-\beta_i')(V_i^0)'
 -S_i'-h_i'\{V_i^0(J_i)-V_i^0\}
 -h_i(V_i^0)'(J_i)J_i'-(V_R^0)'.
\]
The difference $\rho(V_i^0)'-S_i'$ is $O(\rho^{-1})$ by \eqref{eq:V-C1}; every remaining term is of that order. Division by a compact interior speed bound proves the second-derivative bound. The right side is continuous on the specified region, proving $C^2$ regularity. No differentiability of the strategic value at the switch has been assumed.
\end{proof}

\begin{proposition}\label{prop:benchmark}
For all sufficiently large finite $\rho$, the benchmark satisfies part (i) of Theorem~\ref{thm:primitive}. In addition, if
\[
 J_c=J^0(1/2,3)=\frac12+\frac{D_*}{2},
\]
then
\begin{align}\label{eq:rep-premium}
 c^*-c_{\rm dir}&=\frac{\gamma}{\rho}+O(\rho^{-2}),&
 \gamma&=3D_*\{A_L(J_c,3)-A_L(1/2,3)\}>0,\\
 m_\rho:=-\partial_tF_{V^0}(\mu_*,0)&=\frac{\alpha}{\rho}+O(\rho^{-2}),&
 \alpha&=3D_*\{G(J_c)-G(1/2)\}>0.\label{eq:slope-expansion}
\end{align}
The purchase successors of a sufficiently small neighborhood of $\mu_*$ are separated from that neighborhood and from the desk switch.
\end{proposition}
\begin{proof}
By Lemma~\ref{lem:geometry} and uniform value bounds, $\Gamma^0(\mu,p)=pD(\mu,p)+O(\rho^{-1})$. Thus every maximizer is at price $3$ and tends to belief $1/2$. The uniform desk expansion and \eqref{eq:I-margin} make $p_L$ strictly optimal throughout $I$ and $p_H$ strictly optimal at $3/4$, for large $\rho$.

At price $3$, $\Abar(1/2)=1/2$ and $\Abar$ increases with belief. Since $D_*<3/200$, for $\mu\geq1/2$ the jump satisfies
\[
 J^0(\mu,3)-\mu=\frac{\mu(1-\mu)D(\mu,3)}{\Abar(\mu,3)}\leq D_*/2<3/400.
\]
Hence $1/2$, $J_c$, and $J^0(J_c,3)$ lie strictly inside $I$; the last is less than $103/200$. By continuity there is a fixed neighborhood $O$ of $1/2$ with $O,J^0(O,3),J^0(J^0(O,3),3)\subset\operatorname{int}I$. Lemma~\ref{lem:regularity} then gives $C^2$ regularity of the gain on $O$ and
\[
 \Gamma^0_{\mu\mu}(\mu,3)=3D_{\mu\mu}(\mu,3)+O(\rho^{-1}).
\]
Shrink $O$ so that $D_{\mu\mu}$ is uniformly negative. The gain is strictly concave on $O$, and all its global maximizers are there for large $\rho$. There is exactly one, with a strict quadratic gap. Its first-order condition and the derivative expansion give $\mu_*=1/2+O(\rho^{-1})$. Compactness gives a uniform positive gap away from a fixed neighborhood of this state.

For completeness the global reset comparison can be bounded directly. One has $|R_i'|\leq29/5$, $|\beta_i|\leq1/40$, and total event rate at most $2$. For any desk policy, before the first event the reward changes by at most $(29/200)t$, while the event probability is at most $2t$. Since rewards lie in $[0,3]$,
\[
 |\rho\Uti_i^0-R_i|\leq C_0/\rho,\qquad C_0=1229/200.
\]
Also $0\leq V^0\leq3/\rho$, so $|\Gamma^0-pD|\leq\varepsilon_\rho=3D_*/\rho$. Put $\eta_I=D_*-D(13/25,3)>0$ and $g_U=R_H(3/4)-R_L(3/4)>0$. If
\[
 \rho>\max\left\{\frac{2C_0}{\epsilon_I},\frac{2C_0}{g_U},12,\frac{2D_*}{\eta_I}\right\},
\]
every reset to $3$ lies outside $I$, and
\begin{equation}\label{eq:strictwedge3}
 c^*-c^*_{\rm reset}\geq\min\{D_*/2,3\eta_I\}-2\varepsilon_\rho>0.
\end{equation}
This includes all tied reset maximizers. A silence path at price $3$ from $3/4$ to $\mu_*$ has finite travel time and a strictly positive probability of no purchase and no reset.

To establish gain descent, write
\[
 \mathcal J(\mu,\lambda)=\frac{\mu A_H(\lambda,3)}{A_L(\lambda,3)+\lambda D(\lambda,3)},
 \quad F_V(\mu,t)=D(\lambda,3)\{3+V(\mathcal J(\mu,\lambda),3)-V(\mu,3)\}-c,
\]
where $\lambda=\mu+(1-\mu)t$. At a stationary benchmark gain,
\begin{equation}\label{eq:slopeidentity3}
 \partial_tF_{V^0}(\mu,0)
 =-\frac{1-\mu}{\mu}D(\mu,3)
 \{J^0(\mu,3)(V_H^0)'(J^0(\mu,3))-\mu(V_H^0)'(\mu)\}.
\end{equation}
Indeed, along the benchmark $dJ^0/d\mu=J^0/\mu+\mathcal J_\lambda$. Subtract the belief derivative of $\Gamma^0$ from $F_t/(1-\mu)$ and use stationarity. The value is held fixed throughout this calculation.

By \eqref{eq:V-C1}, the expression in braces equals
\[
 \frac3\rho\{G(J^0(\mu_*,3))-G(\mu_*)\}+O(\rho^{-2}).
\]
Both arguments lie in $I$, and their difference converges to $D_*/2>0$. Inequality \eqref{eq:weightedprimitive} makes this expression strictly positive for large $\rho$. This proves strict descent and, using $\mu_*=1/2+O(\rho^{-1})$, \eqref{eq:slope-expansion}.

Finally, \eqref{eq:V-C1}'s level estimate gives
\[
 V_H^0(J_c)-V_H^0(1/2)=\frac3\rho\{A_L(J_c,3)-A_L(1/2,3)\}+O(\rho^{-2})>0.
\]
The function $A_L$ is strictly increasing. Evaluating the gain at its displaced maximum changes the direct peak by only $O(\rho^{-2})$ and the continuation term by that order as well. This proves \eqref{eq:rep-premium}. Since $J^0(\mu_*,3)>\mu_*$ and both states lie strictly below the switch, their small neighborhoods can be chosen disjoint, with all small-trial successors in the latter neighborhood.
\end{proof}

Equations \eqref{eq:rep-premium} and \eqref{eq:slope-expansion} also clarify the order of limits. The cost interval requiring reputation and the negative slope shrink as $\rho$ increases. The local branch is constructed after fixing a finite $\rho$; its estimates are not uniform in a simultaneous limit $\rho\to\infty$, $\delta\to0$.

\section{Localized feedback and exact equilibrium}\label{sec:existence}

Fix a finite $\rho$ large enough for Proposition~\ref{prop:benchmark}. All constants in this section may depend on that fixed $\rho$. Choose a small neighborhood $W$ of $\mu_*$, a small trial bound $t_0>0$, and a compact interval $K_J$ such that:
\begin{enumerate}[label=(\alph*)]
\item $W$ and $K_J$ are disjoint and lie strictly inside the region where the benchmark desk selects $p_L$;
\item every trial successor $\mathcal J(\mu,\mu+(1-\mu)t)$, $\mu\in W$, $0\leq t\leq t_0$, lies in $\operatorname{int}K_J$;
\item for some $m_0>0$, $\partial_tF_{V^0}\leq-2m_0$ on this product, and for some $k>0$,
$c^*-\Gamma^0(\mu,p_H)\geq k(\mu-\mu_*)^2$ on $W$.
\end{enumerate}
These choices follow from the strict slope, curvature, and successor separation already proved. There is also a fixed positive gain gap outside $W\times\{p_H\}$.

\subsection{Value regularity on the actual candidate class}

Consider continuous Lipschitz assessments supported in $W$ at price $p_H$, zero at $p_L$, and uniformly below a fixed $\bar a<1$. The cost lies in a fixed compact neighborhood of $c^*$. For each such assessment the normalized desk operator is a contraction on continuous functions, with continuation mass at most $2/(\rho+2)$. Its fixed point $\Uti^a$ is continuous on $[0,1]\times P$.

On any fixed compact interior set, $|\beta^a|$ has a positive lower bound. The Bellman equation \eqref{eq:UHJB3} therefore gives a common bound on $|(\Uti^a)_\mu|$, from the common value and flow-payoff bounds. This bound does \emph{not} involve $\Lip(a)$: no derivative of $a$ appears in that equation. In particular there is no circular requirement that $\rho$ exceed a bound involving a response Lipschitz constant yet to be chosen.

For a measurable reset kernel that agrees with the benchmark outside a fixed compact switch neighborhood, the strategic next-event operator also preserves continuity. To see this, decompose the reset kernel into a continuous interpolation and a bounded measurable difference supported in that neighborhood. On this compact set, changing variables from silence time to belief integrates the measurable difference against an $L^1$ density that varies continuously in the initial state. All remaining terms are continuous by dominated convergence. The contraction consequently has a continuous fixed point $V^{a,r}$, including at the endpoints. Its Bellman equation holds in the absolutely continuous sense on interior compact intervals and supplies a common local Lipschitz bound, even for highly irregular measurable tie kernels. The bound again depends on values and the minimum silence speed, not on derivatives of the kernel or of $a$. Where $a$ and the reset rule are constant, the right side is continuous and the derivative is classical.

Endpoint assessments remain zero. Algebraic endpoint comparisons are therefore unchanged: $p_L$ is the unique desk winner at $0$ and $p_H$ at $1$. For example, at either endpoint $i\in\{0,1\}$,
\[
 \Uti^a(i,p)=\frac{p b_i(p)+R_i^{\max}/\rho}{\rho+1},\quad
 b_0(p)=\Lambda(-p),\quad b_1(p)=\Lambda(6-p),\quad
 R_i^{\max}=\max_{p'}p'b_i(p').
\]
At $0$, $p\Lambda(-p)$ decreases for $p\geq5/2$; at $1$, the signalwise price comparison in Lemma~\ref{lem:geometry} selects $3$. These statements concern the normalized value; subtracting $c i/\rho$ leaves the ranking unchanged.

\subsection{Occupation and localized value estimates}

For any nearby assessment and either the public-mixture or strategic law, public-event intensity is at most $h=\nu+\kappa=2$. Let a measurable source $f$ be supported in a fixed compact interior set on which every silence speed is at least $b>0$. Partition a path at purchases and reset opportunities, including resets that leave price unchanged. On a segment starting at $T_n$, monotonicity gives
\[
 \int_{T_n}^{T_{n+1}}e^{-\rho t}|f(\mu_t,p_t)|dt
 \leq\frac{e^{-\rho T_n}}b\max_p\int_0^1|f(\mu,p)|d\mu.
\]
Compensation yields $\E\sum_{n\geq0}e^{-\rho T_n}\leq1+h/\rho$. Therefore the positive discounted resolvent satisfies
\begin{equation}\label{eq:occupation3}
 \|\mathcal R f\|_\infty\leq\frac{1+h/\rho}{b}\|f\|_{1,P}.
\end{equation}
There is no division by the vanishing endpoint drift: the source is supported in the interior.

For two assessments $a,b$, changes in flow payoff, drift, hazard, and jump location are bounded pointwise by $C|a-b|$. The local Lipschitz bounds just established, including at all possible purchase successors of the source region, bound the corresponding generator source by the same quantity. Comparing the two optimal aggregate values through their respective exact maximizing kernels and using the two Bellman inequalities gives
\begin{equation}\label{eq:U-stable3}
 \|\Uti^a-\Uti^b\|_\infty\leq C\|a-b\|_{1,P}.
\end{equation}
This is a performance comparison for the literal finite-price maximum, not for a frozen desk selector.

For evaluated strategic values the resolvent identity is
\[
 V^{a,r}-V^{b,s}=\mathcal R^{a,r}
 \big[(g^{a}-g^{b})+(\mathcal L^{a,r}-\mathcal L^{b,s})V^{b,s}\big].
\]
The reset source is bounded by $2\kappa\bar V\ind_{B(r,s)}$, where $B(r,s)$ is the set of beliefs on which some row of the kernels differs. Applying \eqref{eq:occupation3} gives
\begin{equation}\label{eq:V-stable3}
 \|V^{a,r}-V^{b,s}\|_\infty
 \leq C\{\|a-b\|_{1,P}+|B(r,s)|\}.
\end{equation}
The generator identities are mild identities along paths. Almost-everywhere derivatives of locally Lipschitz values suffice, because interior silence occupation is absolutely continuous in belief.

\subsection{A local response lemma}

For small $\delta$, let candidate assessments obey
\begin{equation}\label{eq:envelope3}
 a(\mu,p_L)=0,\qquad
 0\leq a(\mu,p_H)\leq h_\delta(\mu)
 :=\frac{\pos{2\delta-k(\mu-\mu_*)^2}}{m_0}.
\end{equation}
Their support $I_\delta$ is contained in $W$ and their $L^1$ mass is at most
\[
 \int h_\delta=\frac{8\sqrt2}{3m_0\sqrt k}\delta^{3/2}.
\]
Equation \eqref{eq:U-stable3} and the simple benchmark switch imply that every exact candidate desk agrees with the benchmark outside a single ambiguity band $B_\delta$ about $\tau_\rho$, of length $O(\delta^{3/2})$. This follows from the nonzero derivative of the benchmark price difference; no derivative or uniqueness of a perturbed switch is needed. Choose this band to contain every possible candidate tie and to be disjoint from $W$ and $K_J$.

Let $\mathcal R_\delta$ consist of all measurable $2\times2$ stochastic reset kernels that agree with the benchmark outside $B_\delta$, with a row for each old price and the fixed algebraic actions at the endpoints. Within the band either price may receive weight. These kernels need not maximise the candidate desk value. If $r\in\mathcal R_\delta$, then $B(r,r^0)\subseteq B_\delta$ up to a null set. Consequently \eqref{eq:V-stable3} gives
\[
 \|V^{a,r}-V^0\|_\infty
 \leq C\{\|a\|_{1,P}+|B_\delta|\}=O(\delta^{3/2}).
\]
Together with \eqref{eq:U-stable3}, this proves, uniformly over every candidate assessment and every $r\in\mathcal R_\delta$,
\begin{equation}\label{eq:feedbacklevel3}
 \|V^{a,r}-V^0\|_\infty+\|\Uti^a-\Uti^0\|_\infty=O(\delta^{3/2}).
\end{equation}

\begin{lemma}\label{lem:feedback}
For sufficiently small $\delta$, the zero-or-root response to every candidate \eqref{eq:envelope3} and every $r\in\mathcal R_\delta$ is continuous, has a common Lipschitz bound in belief, is bounded above by $h_\delta$, and has trial slope at most $-m_0$ on $W\times[0,t_0]$. It is zero at the other price and off $I_\delta$.
\end{lemma}
\begin{proof}
On the separated successor interval $K_J$, every candidate assessment is zero and every $r\in\mathcal R_\delta$ resets to $p_L$, because $K_J$ lies outside $B_\delta$. Subtract the strategic Bellman equations and put $\Delta V=V^{a,r}-V^0$. At either current price,
\[
 \beta_i^0\Delta V_i'=(\rho+A_{L,i}^0+1)\Delta V_i
 -A_{L,i}^0\Delta V_i(J_i^0)-\Delta V_L.
\]
Hence, with $b_J=\min_{K_J,i}|\beta_i^0|>0$,
\begin{equation}\label{eq:derivativefeedback3}
 \|\Delta V'\|_{\infty,K_J}\leq\frac{\rho+4}{b_J}\|\Delta V\|_\infty
 =O(\delta^{3/2}).
\end{equation}
The derivative is classical here. In $\partial_tF_V$, the only derivative of $V$ is at the trial purchase successor; the value at the current belief enters only in levels. Equations \eqref{eq:feedbacklevel3}--\eqref{eq:derivativefeedback3} therefore make the trial slope uniformly close to the baseline slope, proving the asserted bound.

At zero trial,
\[
 F_{V^{a,r}}(\mu,0)\leq\delta+C\delta^{3/2}-k(\mu-\mu_*)^2
 \quad(\mu\in W).
\]
For small $\delta$, this is at most $m_0h_\delta$. Set the response to zero when this gain is nonpositive; otherwise strict descent gives a unique root in $(0,h_\delta]\subset(0,t_0]$. The fixed exterior and other-price gain gaps make quality zero everywhere else. The response need not search for other roots outside this small-trial interval.

On $W$, the common local Lipschitz bounds for evaluated values and the smooth primitive trial maps give a bound $C_F$ on the belief Lipschitz modulus of $F_V$, uniform in candidates and desks. Strict descent makes the zero-or-root response Lipschitz with modulus at most $C_F/m_0$. The response vanishes in a neighborhood of the boundary of $W$, so extension by zero is globally Lipschitz. The constants were obtained without using the candidate's own Lipschitz modulus.
\end{proof}

\subsection{Desk graph, continuity, and fixed point}

Choose a finite $L$ larger than the response modulus in Lemma~\ref{lem:feedback}, and let $\mathcal A_\delta$ be the assessments satisfying \eqref{eq:envelope3} and $\Lip(a)\leq L$. This set is nonempty, convex, and compact in the uniform norm by Arzel\`a--Ascoli. The preceding regularity and source bounds are uniform on this class.

Endow the entries of $\mathcal R_\delta$ with the weak-star topology $\sigma(L^\infty,L^1)$, identifying kernels that agree almost everywhere in the interior. Positivity, row sums, and support restrictions are weak-star closed. The set is convex and compact by Banach--Alaoglu, and metrizable because $L^1[0,1]$ is separable. The fixed endpoint actions are imposed separately and do not vary in this topology.

For each $a$, let $M^a(\mu)=\max_j\Uti_j^a(\mu)$ and define
\[
 \mathcal D(a)=\{r\in\mathcal R_\delta:
 r_{ij}(\mu)[M^a(\mu)-\Uti_j^a(\mu)]=0\ \hbox{a.e., for all }i,j\}.
\]
The correspondence has nonempty compact convex values; a least-index exact maximizer is a Borel selection. Its graph is closed. Indeed, if $a_n\to a$ uniformly and $r_n\rightharpoonup^*r$ with $r_n\in\mathcal D(a_n)$, the value-gap functions converge uniformly by \eqref{eq:U-stable3}. Testing the displayed products against arbitrary $L^1$ functions passes exact support to the limit.

We also need continuity of the strategic value under weak-star convergence of kernels, rather than pointwise convergence of selectors. Before the next event, the density with which a reset tests the kernel at an interior belief $y$ is
\begin{equation}\label{eq:density3}
 k_{\mu,i}^{a}(y)=\frac{\kappa\ind_{\{y\text{ on downstream orbit}\}}}{-\beta_i^a(y)}
 \exp\!\left[-\rho T_i^a(\mu,y)-\int_0^{T_i^a(\mu,y)}
 \{\kappa+\nu A_{S,i}^a(\phi_i^a(s,\mu))\}\,ds\right].
\end{equation}
Restrict this density to $B_\delta$. The denominator has a common positive lower bound. With the common candidate Lipschitz bound, flows and hazards depend continuously on assessments and initial states on finite time horizons. Changing the orbit indicator moves only an interval endpoint. It follows by dominated convergence that these restricted densities vary in $L^1$; the discounted tail is uniformly negligible. For initial beliefs below the band the density is zero. As the initial belief tends to one, the time to the band tends to infinity and the density norm tends to zero. Compactness of the assessment and initial-state sets therefore makes the family norm-compact in $L^1(B_\delta)$.

For this particular construction there is a further simplification: $B_\delta$ lies strictly above the effort region. A decreasing no-event path reaching the band has not yet visited that region. The density on $B_\delta$ is consequently the zero-quality density, independent of the candidate assessment. Only its initial-state argument varies.

A bounded weak-star convergent sequence converges uniformly when tested against a norm-compact subset of $L^1$, by a finite-net argument. Apply this to \eqref{eq:density3} multiplied by a fixed continuous continuation value. Outside $B_\delta$ the kernels coincide. The other next-event terms are continuous by the uniform flow convergence and dominated convergence. Thus
\[
 \|(T^V_{a_n,r_n}-T^V_{a,r})w\|_\infty\longrightarrow0
 \quad\hbox{for each continuous }w,
\]
and the common sup-norm contraction implies $\|V^{a_n,r_n}-V^{a,r}\|_\infty\to0$.

Set $X_\delta=\mathcal A_\delta\times\mathcal R_\delta$. Lemma~\ref{lem:feedback} defines the genuine zero-or-root response $Q(a,r)$ on this entire product, including kernels that do not maximise the candidate desk value. Uniform convergence of values gives uniform convergence of trial gains. The common strictly negative slope makes roots converge uniformly, including at zero-root boundaries. Thus $Q:X_\delta\to\mathcal A_\delta$ is continuous.

The product $X_\delta$ is nonempty, compact, metrizable, and convex in the Hausdorff locally convex space obtained from the uniform topology for assessments and the weak-star topology for kernels. Define
\[
 \Phi(a,r)=\{Q(a,r)\}\times\mathcal D(a).
\]
This correspondence has nonempty compact convex values and a closed graph. The fixed-point theorem of \citet{Glicksberg1952}, p.~171, therefore applies directly. At a fixed point $(a_\delta,r_\delta)$, the first coordinate satisfies the genuine quality response and the second is an exact desk maximiser: $a_\delta=Q(a_\delta,r_\delta)$ and $r_\delta\in\mathcal D(a_\delta)$. No extension or convexification of the quality response is needed.

At $\mu_*$ the zero-trial gain is at least $\delta-C\delta^{3/2}>0$, so quality is strictly positive. The envelope proves its height, width, and mass bounds, and \eqref{eq:feedbacklevel3} proves the value bounds. Exact reset support initially holds almost everywhere. Modify the kernel on the exceptional null set to select a Borel exact maximizer. Nonzero interior silence speed makes reset-belief occupation absolutely continuous. Even if a purchase lands at an exceptional point, the next reset occurs strictly later almost surely and silence leaves the point immediately. This modification changes neither values nor effort, while giving pointwise desk obedience. Endpoint actions are already exact.

The strategic Bellman equation evaluated at the response is the maximizing hidden-action equation: high and low quality have the same value when mixing, and low quality is optimal at zero response. Integrating that Bellman inequality against any alternative private quality strategy and using bounded discounted continuation values verifies optimality, including history-dependent deviations. The filter is Bayesian under the constructed assessment, by \eqref{eq:filter3}. This establishes stationary equilibrium in the original model.

Finally, the low-price margin on the effort region and the high-price margin at $3/4$ persist by \eqref{eq:U-stable3}. The sale-free, reset-free upstream path has strictly positive probability. This completes Theorem~\ref{thm:primitive}. Corollary~\ref{cor:essential} follows from $c>c_{\rm dir}=\max_{\lambda,p}pD(\lambda,p)$ and the equality condition for positive mixed quality.

\section{Local adjustment and proofs of learning}\label{sec:dynamicproofs}

The local-adjustment statement is Theorem~\ref{thm:pulse} in Section~\ref{sec:dynamics}. We first prove learning and the lifetime upper bounds, then establish the local profile and the matching lifetime lower bounds.

\subsection{Learning and permanent endpoint-price selection}

Under the public mixture law, the Bayesian posterior is a bounded martingale with predictable quadratic variation
\[
 \langle\mu\rangle_t=\int_0^t\nu\Abar_s(J_s-\mu_s)^2ds.
\]
If $a\leq\bar a<1$, then for every compact interval $K\subset(0,1)$,
\begin{equation}\label{eq:mK3}
 m_K:=\min_{\mu\in K,\ p\in P,\ 0\leq t\leq\bar a}
 \nu\Abar(\mu,p,t)\{J(\mu,p,t)-\mu\}^2>0.
\end{equation}
Indeed, $J-\mu=\mu(1-\mu)(1-t)D/\Abar>0$ on this compact set. Since
$\E\langle\mu\rangle_\infty\leq\mu_0(1-\mu_0)$,
\begin{equation}\label{eq:undiscK3}
 \E\int_0^\infty\ind_K(\mu_s)ds\leq\frac{\mu_0(1-\mu_0)}{m_K},\qquad
 \E_S\int_0^\infty\ind_K(\mu_s)ds\leq\frac{\mu_0}{m_K}.
\end{equation}
The posterior converges almost surely. An interior limit would imply infinite occupation of some compact interior interval, contradicting \eqref{eq:undiscK3}. Thus its limit is in $\{0,1\}$. Bayesian consistency gives $\mu_\infty=\E[\ind_{\{\theta=C\}}\mid\mathcal F_\infty]$, and therefore
\[
 \E(\ind_{\{\theta=C\}}-\mu_\infty)^2=\E\mu_\infty(1-\mu_\infty)=0.
\]
This identifies the boundary with the true type.

The branch has effort supported in a fixed compact interior region and $a_\delta\leq\bar a<1$ for small $\delta$. Posterior convergence implies a finite last time in that region. Endpoint continuity of $\Uti_\delta$ was established above; the strict algebraic endpoint winners therefore remain unique on endpoint neighborhoods. Eventually belief remains in the appropriate neighborhood, and the next rate-$\kappa$ reset implements its winner permanently. This proves the learning, cessation, and price-selection assertions of Theorem~\ref{thm:learning}.

\subsection{Undiscounted lifetime quality}

Let $E_\delta=[\ell_\delta,u_\delta]$ be the hull of the effort support, within a fixed $[\ell,u]\subset(0,1)$. Uniformly for small $\delta$, its width is $O(\sqrt\delta)$ and all silence speeds on it exceed some $b>0$. Put $b_{\min}=\min_p\Lambda(-p)>0$. The purchase map obeys $J\leq\mu/b_{\min}$. Any purchase beginning below the upper edge and landing in or above the band must therefore start in the fixed compact interval
$K=[\ell b_{\min},u]$. Purchases beginning in the band also start in $K$.

Let $T_E(T)$ and $T_K(T)$ be occupations up to deterministic time $T$. Each uninterrupted downward traversal contributes at most $|E_\delta|/b$. Beyond an initial traversal, a further piece of occupation requires a purchase from $K$ or a reset within $E_\delta$. Resets above or below the band do not create a new crossing by themselves, since they do not change belief. Counting these events, and then taking expectations using their intensities, gives
\[
 \E_S T_E(T)\leq\frac{|E_\delta|}{b}
 \{1+\nu\E_S T_K(T)+\kappa\E_S T_E(T)\}.
\]
This truncated argument avoids subtracting infinite expectations. For $\kappa|E_\delta|<b$, rearrange and let $T\to\infty$. Equation \eqref{eq:undiscK3}, with a common $\bar a<1$, yields
\begin{equation}\label{eq:thinundisc3}
 \E_S T_E\leq\frac{|E_\delta|}{b-\kappa|E_\delta|}
 \left(1+\frac{\nu\mu_0}{m_K}\right)=O(\sqrt\delta).
\end{equation}
The private opportunity clock has intensity $\nu$, and a high-quality decision has conditional probability $a_\delta$. Its marked compensator gives
\[
 \E_SN_H=\nu\E_S\int_0^\infty a_\delta(\mu_t,p_t)dt
 \leq\nu\|a_\delta\|_\infty\E_ST_E=O(\delta^{3/2}).
\]
Markov's inequality proves the probability upper bound. The positive matching lower bounds from the upstream state follow after the local calculation below.

\subsection{Local profile and first traversal}

The baseline gain is $C^2$ at its unique peak and its trial map is $C^1$ near $(\mu_*,0)$. The value-level error is $O(\delta^{3/2})$, uniformly over the equilibria constructed above. For $t=O(\delta)$ and bounded $y$ this gives
\[
 F_{V_\delta}(\mu_*+\sqrt\delta\,y,t)
 =\delta\left(1-\frac K2y^2\right)-mt+o(\delta),
\]
uniformly. The zero-or-root condition and the uniform negative slope imply \eqref{eq:pulse3}. The same sign comparison, together with the quadratic gap and global support bound, confines the support edges to $\mu_*\pm Y\sqrt\delta+o(\sqrt\delta)$. It does not assume that the perturbed support is exactly one interval; its positive core and its hull have these limits.

Along the no-event path from the upper edge to the lower edge, $-\beta^\delta\to b_*$ uniformly. Changing from time to belief therefore gives $T_\delta=(2Y/b_*)\sqrt\delta+o(\sqrt\delta)$. The jump size converges to a strictly positive number while the support width vanishes, so a purchase exits above the upper edge. Throughout this region the unique reset price is $p_L$.

Under the strategic-type law, the probability of surviving without a public event to time $t$ along this path is
\[
 \mathcal S_\delta(t)=\exp\!\left[-\kappa t-\int_0^t\nu A_S^\delta(\mu_s,p_H)ds\right].
\]
Its purchase hazard converges uniformly to $h_S^*$, and $T_\delta=O(\sqrt\delta)$. Integrating $\kappa\mathcal S_\delta$ and $\nu A_S^\delta\mathcal S_\delta$ up to $T_\delta$, and evaluating $\mathcal S_\delta(T_\delta)$, proves the three exit probabilities.

The expected number of high-quality decisions before or at termination of this traversal is
\[
 \nu\int_0^{T_\delta}\mathcal S_\delta(t)a_\delta(\mu_t,p_H)dt.
\]
This expression includes the opportunity generating a terminating purchase. Using the profile, changing to rescaled belief, and noting $\mathcal S_\delta=1+O(\sqrt\delta)$ gives
\[
 \frac{\nu\delta^{3/2}}{mb_*}\int_{-Y}^Y\left(1-\frac K2y^2\right)dy
 +o(\delta^{3/2})
 =\frac{4\nu Y}{3mb_*}\delta^{3/2}+o(\delta^{3/2}).
\]
This proves Theorem~\ref{thm:pulse}.

From $(3/4,p_H)$, the strategic-type probability of reaching the upper edge with neither a purchase nor a reset converges to a strictly positive constant. Up to entry the assessment is zero, so the path and hazards are the benchmark ones, and its limiting travel time is finite. Multiplying by \eqref{eq:pulsecount3} proves the lifetime expectation lower bound of order $\delta^{3/2}$.

For the probability lower bound, on the first traversal high-quality marks are dominated by a rate-$\nu\|a_\delta\|_\infty$ Poisson count over the deterministic time $T_\delta$. If $N^{\rm tr}$ is the actual count, then
\[
 \E[N^{\rm tr}(N^{\rm tr}-1)]
 \leq(\nu\|a_\delta\|_\infty T_\delta)^2=O(\delta^3).
\]
For a nonnegative integer count, $\Prb(N\geq1)\geq\E N-\tfrac12\E[N(N-1)]$. Hence the chance of at least one high-quality decision on that traversal is also of order $\delta^{3/2}$. The positive entry probability and the lifetime upper bound finish Theorem~\ref{thm:learning}.

% END Proofs.tex

% BEGIN Robustness_Proof.tex
\section{Robustness on an open set of primitives}\label{sec:robustproof}

We prove Proposition~\ref{prop:robustness}. Prices are identified by their labels, so all value functions have the common domain $[0,1]\times\{L,H\}$ even when the two price levels vary. A subscript $\xi$ denotes the zero-quality economy at that parameter vector. The discount rate is one coordinate of $\xi$ and remains in a compact neighbourhood of the fixed finite $\rho_0$. All neighbourhoods below are chosen within the admissible parameter space.

\subsection{Continuity of the benchmark values and the moving switch}

Begin with a compact parameter neighbourhood of $\xi_0$. The purchase probabilities, jump maps and silence vector fields at zero assessed quality are jointly smooth in belief and parameters. Their denominators are bounded away from zero. In particular, with
\[
 \underline\rho=\min_\xi\rho>0,\qquad
 \overline h=\max_\xi(\nu+\kappa),\qquad
 b_{\min}=\min_{\xi,i}\Lambda(-p_i/\zeta)>0,
\]
the total event intensity is at most $\overline h$, and the first-event operators have a common contraction modulus no greater than
\[
 \chi=\frac{\overline h}{\underline\rho+\overline h}<1.
\]
On every fixed compact belief interval in $(0,1)$, the magnitude of the zero-quality silence drift also has a common strictly positive lower bound.

The optimal desk operator is \eqref{eq:firstU} with the relevant rates restored: its flow reward is $\nu p_i\Abar_{\xi,i}^0$, its purchase rate is $\nu\Abar_{\xi,i}^0$, and its reset term is $\kappa\max_j w_j$. For each fixed continuous $w$, this operator depends continuously on $\xi$ in the uniform norm. To see this, truncate the integral at $T$. Smooth vector fields imply uniform convergence of the flows on $[0,T]$, uniformly over initial beliefs. All reward, survival and continuation terms then converge uniformly. The remaining integral is uniformly bounded by a constant times $e^{-\underline\rho T}$. Taking $T$ large proves the assertion. The common contraction yields
\begin{equation}\label{eq:robUcontinuity}
 \|\Uti_\xi^0-\Uti_{\xi'}^0\|_\infty\longrightarrow0
 \qquad\hbox{as }\xi'\to\xi.
\end{equation}

This argument is for the optimal desk operator, including its price maximum. It does not freeze a selector. In the interior, its Bellman equation gives
\begin{equation}\label{eq:robUderivative}
 \beta_{\xi,i}^0(\Uti_{\xi,i}^0)'
 =(\rho+\nu\Abar_{\xi,i}^0+\kappa)\Uti_{\xi,i}^0
 -\nu p_i\Abar_{\xi,i}^0
 -\nu\Abar_{\xi,i}^0\Uti_{\xi,i}^0(J_{\xi,i}^0)
 -\kappa\max_j\Uti_{\xi,j}^0.
\end{equation}
The right side is continuous. The integral equation along a nonzero silence flow therefore gives a classical first derivative locally. Dividing by the drift shows that \eqref{eq:robUcontinuity} strengthens to $C^1$ convergence on compact interior intervals.

At $\xi_0$, the desk gap $\Uti_H^0-\Uti_L^0$ has one zero with positive derivative. Choose a small closed interval around that zero where the derivative remains strictly positive. On its compact complement the gap is bounded away from zero. Uniform convergence of levels preserves these exterior signs, and $C^1$ convergence preserves the positive derivative on the interval. Every sufficiently nearby economy thus has exactly one simple switch $\tau(\xi)$, continuous in $\xi$. At the switch both prices are optimal; either choice has zero reset-time occupation and gives the same benchmark values.

Next consider the strategic value evaluated at this zero-quality desk. Write $r_\xi^0(\mu)=L$ below $\tau(\xi)$ and $H$ above it. Continuity of this value requires an additional argument, since $r_\xi^0$ need not converge uniformly. Fix a continuous continuation function $w$ and truncate the strategic first-event operator at $T$. If $\xi_n\to\xi$, let $\varepsilon_n$ be the maximum distance between the two no-event flows up to $T$; then $\varepsilon_n\to0$. The reset-price indicators can disagree only while the limiting flow lies within distance
\[
 \varepsilon_n+|\tau(\xi_n)-\tau(\xi)|
\]
of the limiting switch. This interval lies in a fixed compact interior set. Monotonicity and a common lower bound $b$ on silence speed bound the time spent there by twice its radius divided by $b$, uniformly over initial beliefs. The resulting reset-integral difference tends to zero. All other truncated terms converge by smoothness and uniform continuity of $w$; the discounted tails are uniformly negligible. The common contraction now gives
\begin{equation}\label{eq:robVcontinuity}
 \|V_\xi^0-V_{\xi'}^0\|_\infty\longrightarrow0.
\end{equation}
Thus movement of the endogenous switch has been accounted for explicitly.

For a fixed compact interval disjoint from the nearby switches, the reset label is locally constant. Put $h_{\xi,i}=\nu A_{L,\xi,i}^0$ and $g_{\xi,i}=\nu p_i A_{L,\xi,i}^0$. The strategic Bellman equation gives
\begin{equation}\label{eq:robVderivative}
 \beta_{\xi,i}^0(V_{\xi,i}^0)'
 =(\rho+h_{\xi,i}+\kappa)V_{\xi,i}^0-g_{\xi,i}
 -h_{\xi,i}V_{\xi,i}^0(J_{\xi,i}^0)
 -\kappa V_{\xi,r_\xi^0}^0.
\end{equation}
It follows that the derivatives converge uniformly on this interval. If its purchase image also avoids the switches, differentiating once more gives
\begin{align}\label{eq:robVsecond}
 \beta_{\xi,i}^0(V_{\xi,i}^0)''
 ={}&(\rho+h_{\xi,i}+\kappa-(\beta_{\xi,i}^0)')(V_{\xi,i}^0)'
 -g_{\xi,i}'\notag\\
 &-h_{\xi,i}'\{V_{\xi,i}^0(J_{\xi,i}^0)-V_{\xi,i}^0\}
 -h_{\xi,i}(V_{\xi,i}^0)'(J_{\xi,i}^0)(J_{\xi,i}^0)'
 -\kappa(V_{\xi,r_\xi^0}^0)'.
\end{align}
All terms converge uniformly on such separated compact sets. These identities establish the derivative control needed below, without asserting differentiability through the strategic-value kink at the switch.

\subsection{Persistence of the strict geometry}

Choose a compact interval $W$ around the baseline gain maximum. As in Appendix~\ref{sec:baselineproof}, its high-price purchase image and that image's purchase image lie strictly below the baseline switch; $W$ and its first image are disjoint. After shrinking both $W$ and the parameter neighbourhood, these separations hold for every nearby economy, and all small-trial purchase successors from $W$ lie in a fixed compact interval $K_J$ disjoint from $W$ and the switches.

The benchmark gain
\[
 \Gamma_\xi^0(\mu,p_i)
 =D_\xi(\mu,p_i)\{p_i+V_{\xi,i}^0(J_{\xi,i}^0(\mu))-V_{\xi,i}^0(\mu)\}
\]
is uniformly continuous in parameters on the whole state space. Equations \eqref{eq:robVderivative}--\eqref{eq:robVsecond}, applied on $W$ and its first two high-price purchase images, give $C^2$ continuity of its high-price component on $W$. The strict exterior gain gap at the baseline confines every nearby global maximizer to $W\times\{H\}$. Strict negative curvature on $W$ then gives a unique maximum $\mu_*(\xi)$ with $\mu_*(\xi)\to\mu_*(\xi_0)$. Its height $c^*(\xi)$ and curvature
\[
 K(\xi)=-\partial_{\mu\mu}\Gamma_\xi^0(\mu_*(\xi),p_H)>0
\]
are continuous. The maximum stays in the interior of $W$.

For trial quality $t$, the assessment is $\lambda=\mu+(1-\mu)t$ and the purchase successor is
\[
 \mathcal J_\xi(\mu,\lambda)
 =\frac{\mu A_{H,\xi}(\lambda,p_H)}
 {A_{L,\xi}(\lambda,p_H)+\lambda D_\xi(\lambda,p_H)}.
\]
The trial gain and its $t$ derivative use only levels of $V_\xi^0$ at the current belief and the first derivative at this successor. Hence \eqref{eq:robVcontinuity}--\eqref{eq:robVderivative} imply joint continuity on a small product $W\times[0,t_0]$ and the parameter neighbourhood. The baseline strict negative trial slope therefore persists with a common margin. In particular,
\[
 m(\xi)=-\partial_tF_{V_\xi^0,\xi}(\mu_*(\xi),0)>0
\]
is continuous. This calculation holds the continuation function fixed while changing the assessment, as in Theorem~\ref{thm:pulse}.

The immediate-return threshold is continuous by maximization of a jointly continuous function on a fixed compact set. The reset-restricted threshold can be written
\begin{equation}\label{eq:robresetthreshold}
 c^*_{\rm reset}(\xi)=\max\left\{
 \max_{0\leq\mu\leq\tau(\xi)}\Gamma_\xi^0(\mu,p_L),\quad
 \max_{\tau(\xi)\leq\mu\leq1}\Gamma_\xi^0(\mu,p_H)
 \right\}.
\end{equation}
Each moving interval can be parametrized by $[0,1]$, proving continuity. This expression includes both prices at a tie. The two strict threshold gaps at $\xi_0$ therefore remain positive nearby. The strict low-price ranking on $W\cup K_J$, high-price ranking at $3/4$, and endpoint rankings likewise persist.

We may now choose an open $\mathcal N$ with compact closure entirely within this neighbourhood. On its closure there are common positive lower bounds for the threshold gaps, gain curvature, trial descent, switch derivative, exterior gain gap, successor separation and the required interior silence speeds. The endpoints also retain their original strict price rankings. These choices prove part (i) of the proposition.

\subsection{A common cost interval and exact equilibria}

For each $\xi\in\overline{\mathcal N}$, compare candidate assessments with that economy's own zero-quality benchmark. The occupation bound of Appendix~\ref{sec:existence} becomes
\begin{equation}\label{eq:roboccupation}
 \|\mathcal R_\xi f\|_\infty
 \leq\frac{1+\overline h/\underline\rho}{b}\|f\|_{1,P}
\end{equation}
for sources supported on the fixed interior sets under consideration. Candidate quality is uniformly bounded away from one. Common bounds for value levels and the Bellman equations give common local Lipschitz bounds at those sources and their purchase images. Smoothness of the primitive coefficients and boundedness of the nearby costs then give, with one finite constant $C$,
\begin{align}\label{eq:robfeedback}
 \|\Uti_\xi^a-\Uti_\xi^0\|_\infty
 &\leq C\|a\|_{1,P},\notag\\
 \|V_\xi^{a,r}-V_\xi^0\|_\infty
 &\leq C\{\|a\|_{1,P}+|B(r,r_\xi^0)|\}.
\end{align}
The first comparison uses the two optimal desk Bellman inequalities. The second permits any candidate reset kernel differing from the benchmark only in its ambiguity band. Neither estimate differentiates a selected positive-quality equilibrium with respect to parameters.

Choose common $k,m_0>0$ smaller than the curvature and descent margins just established. At cost $c=c^*(\xi)-\delta$, use the envelope
\[
 h_{\xi,\delta}(\mu)
 =m_0^{-1}\pos{2\delta-k(\mu-\mu_*(\xi))^2}
\]
at price $p_H$ and zero at $p_L$. Its height is $O(\delta)$, width $O(\sqrt\delta)$ and mass $O(\delta^{3/2})$, with common constants. The desk's simple benchmark switch and \eqref{eq:robfeedback} confine every candidate desk change to a band $B_{\xi,\delta}$ around $\tau(\xi)$ of length $O(\delta^{3/2})$. This band stays above and separate from $W$ and $K_J$.

For every kernel in the enlarged candidate class, \eqref{eq:robfeedback} gives $O(\delta^{3/2})$ value errors. On $K_J$, candidate quality is zero and the reset label is $L$. Subtracting \eqref{eq:robVderivative} from the candidate equation gives
\[
 \beta_{\xi,i}^0\Delta V_i'
 =(\rho+h_{\xi,i}+\kappa)\Delta V_i
 -h_{\xi,i}\Delta V_i(J_{\xi,i}^0)-\kappa\Delta V_L.
\]
The common lower drift bound implies
\[
 \|\Delta V'\|_{\infty,K_J}
 \leq C\|\Delta V\|_\infty=O(\delta^{3/2}).
\]
The genuine zero-or-root quality response thus has a common negative trial slope, is bounded by $h_{\xi,\delta}$ and has a common belief Lipschitz bound, throughout the product of candidate assessments and enlarged reset kernels. The exterior gain gap sets the response to zero elsewhere.

For each fixed $\xi$ and $\delta$, apply the compact-product construction in Appendix~\ref{sec:existence}: continuous Lipschitz assessments have the uniform topology, and reset kernels have the weak-star topology. The exact desk correspondence has a closed graph. Strategic values are continuous under weak-star kernel convergence because reset densities on the interior ambiguity band form a norm-compact $L^1$ family. The density argument is unchanged when the positive rates and drift are replaced by their nearby values. The quality response is continuous on the entire product. Glicksberg's theorem then gives an exact fixed point for this economy. Modification on null sets and the Bellman verification give pointwise desk obedience and optimality against arbitrary private quality deviations, as before.

All smallness requirements used in the envelope, feedback, source separation and obedience arguments depend only on the common bounds on $\overline{\mathcal N}$. Choose one $\delta_0>0$ satisfying them, and also
\[
 \delta_0<\frac12\min_{\xi\in\overline{\mathcal N}}
 \{c^*(\xi),\ c^*(\xi)-c_{\rm dir}(\xi),\
 c^*(\xi)-c^*_{\rm reset}(\xi)\}.
\]
The quality cost is positive and exceeds every immediate quality return. At the gain maximum the zero-trial gain is $\delta+O(\delta^{3/2})>0$. The strict low-price ranking on the support and high-price ranking at $3/4$ persist after the value perturbation. The upstream no-purchase, no-reset path has positive probability of reaching the support. This proves part (ii). The fixed point is constructed anew for each parameter vector; a unique or continuous equilibrium selection is unnecessary.

\subsection{Learning, adjustment and lifetime quality}

The learning proof in Appendix~\ref{sec:dynamicproofs} is already stated for finite menus and arbitrary positive rates. On a fixed compact interior belief interval, its positive information bound can now be minimized jointly over $\overline{\mathcal N}$ and the common range $0\leq a\leq\bar a<1$. It remains strictly positive. Thus every constructed equilibrium learns the true type, and strategic quality ceases after a finite random time.

To check permanent prices with the rates restored, write $e\in\{0,1\}$ for an endpoint belief and set
\[
 b_0(p)=\Lambda(-p/\zeta),\qquad b_1(p)=\Lambda((v-p)/\zeta).
\]
For any candidate with zero endpoint quality, the normalized desk value there is
\[
 \Uti_\xi^a(e,p_i)
 =\frac{\nu p_i b_e(p_i)+(\kappa\nu/\rho)\max_j p_jb_e(p_j)}{\rho+\kappa}.
\]
The strict endpoint winners persist by continuity of these algebraic comparisons. For each constructed equilibrium, value continuity extends the rankings to endpoint neighbourhoods. Eventual learning and the next price-reset opportunity then give permanent $p_L$ under the strategic type and $p_H$ under the committed type.

Put
\[
 Y(\xi)=\sqrt{2/K(\xi)},\quad
 b_*(\xi)=-\beta_\xi^0(\mu_*(\xi),p_H),\quad
 h_S^*(\xi)=\nu A_{L,\xi}^0(\mu_*(\xi),p_H).
\]
These coefficients and $m(\xi)$ are continuous and strictly positive on the compact parameter set. Continuity in $\xi$ of the first two belief derivatives of the baseline gain, joint continuity of its trial derivative, and the uniform value error give
\begin{equation}\label{eq:robprofileexpansion}
 F_{V_{\xi,\delta},\xi}(\mu_*(\xi)+\sqrt\delta\,y,t)
 =\delta\{1-K(\xi)y^2/2\}-m(\xi)t+o(\delta)
\end{equation}
uniformly in $\xi$, bounded $y$, and $t=O(\delta)$, over every equilibrium furnished by the construction. The zero-or-root rule proves the uniform pulse profile. Its sign on either side of the two limiting roots, together with the common envelope, gives the support-hull endpoints uniformly. This argument makes no connected-support assertion.

The silence speed on the shrinking hull converges uniformly to $b_*(\xi)$. The purchase jump is uniformly separated from zero, so it crosses above the hull for small $\delta$, while a reset selects the inactive low price. The survival and hazard integrals in Appendix~\ref{sec:dynamicproofs} therefore give, uniformly on $\overline{\mathcal N}$,
\[
 T_{\xi,\delta}=\frac{2Y(\xi)}{b_*(\xi)}\sqrt\delta+o(\sqrt\delta),
 \qquad
 \E_S N^{\rm tr}_{\xi,\delta}
 =\frac{4\nu Y(\xi)}{3m(\xi)b_*(\xi)}\delta^{3/2}
 +o(\delta^{3/2}),
\]
as well as the three exit probabilities in Theorem~\ref{thm:pulse} with the corresponding $\kappa$, $h_S^*(\xi)$ and $b_*(\xi)$.

For lifetime counts, place every support hull in a fixed $[\ell,u]\subset(0,1)$. The bound $J\leq\mu/b_{\min}$ puts every purchase capable of creating another traversal in $K=[\ell b_{\min},u]$. With $\overline\nu=\max\nu$, $\overline\kappa=\max\kappa$ and a common information bound $m_K>0$, the undiscounted occupation argument gives
\[
 \E_S T_{E_{\xi,\delta}}
 \leq\frac{|E_{\xi,\delta}|}{b-\overline\kappa|E_{\xi,\delta}|}
 \left(1+\frac{\overline\nu\mu_0}{m_K}\right)
 =O(\sqrt\delta)
\]
uniformly, after decreasing $\delta_0$ if needed. Multiplying by $\nu\|a_{\xi,\delta}\|_\infty$ proves the uniform lifetime expectation bound and hence the probability upper bound.

From $(3/4,p_H)$, the quiet entry path stays in a fixed compact interior interval and has uniformly bounded travel time. Its event intensity is bounded by $\overline h$, so its entry probability has a common positive lower bound. The traversal-count coefficient above also has a positive minimum. The factorial-second-moment bound remains $O(\delta^3)$ uniformly. Consequently both the lifetime expectation and the probability of at least one high-quality decision have matching positive lower bounds of order $\delta^{3/2}$. This proves part (iii) and completes the proposition.

All limits in this appendix send $\delta$ to zero while $\xi$ remains in the fixed compact neighbourhood of $\xi_0$. The neighbourhood may shrink with the baseline discount rate. The proof does not change the demand family, enlarge the price menu, or relax the pricing desk's information restriction.

\begingroup\singlespacing

\endgroup
\end{document}